\documentstyle[12pt]{article}
\def\beqa{\begin{eqnarray}}
\def\eqa{\end{eqnarray}}
\newcommand{\kb}{\mbox{$\underline {k}$}}
\newcommand{\xb}{\mbox{$\underline {x}$}}
\newcommand{\bb}{\mbox{$\underline {b}$}}

\newcommand{\epsilonb}{\mbox{$\underline {\epsilon}$}}

\newcommand{\Rb}{\mbox{$\underline {R}$}}

\newcommand{\wb}{\mbox{$\underline {w}$}}
\newcommand{\qb}{\mbox{$\underline {q}$}}
\newcommand{\de}{\mbox{$\frac{1}{2}$}}

\def\beq{\begin{equation}}
\def\eq{\end{equation}}

\renewenvironment{thebibliography}[1]
          {\begin{list}{[\arabic{enumiv}]}
          {\usecounter{enumiv}\setlength{\parsep}{0pt}
\setlength{\leftmargin .75cm}{\rightmargin 0pt}
        \setlength{\itemsep}{8pt} \settowidth
        {\labelwidth}{#1.}\sloppy}}{\end{list}}
\begin{document}
\input epsf
\parindent 30pt 
\begin{titlepage}
\vspace*{-1.5cm}

\begin{center}
\baselineskip=13pt

{\hspace*{10cm} Saclay-T97/023, DESY 97-046
  }
\vspace{2cm}

{\Large \bf  Onium-Onium scattering at fixed impact parameter: exact
equivalence between the color dipole model and the BFKL Pomeron. \\}
\vskip2.5cm
{\Large Henri Navelet}\\
\vskip1.5cm
{\it Service de Physique Th\'eorique\footnote{Laboratoire de la Direction
des
Sciences de la Mati\`ere du Commissariat \`a l'Energie
Atomique}, CE-Saclay \\ 91191 Gif-sur-Yvette, France}\\
\vskip2.5cm
{\Large Samuel Wallon} \footnote{Alexander von Humboldt Fellow}\\
\vskip1.5cm
{\it II. Institut f\"ur Theoretische Physik, Universit\"at Hamburg \\  Hamburg, Germany}\\
\end{center}
\vspace*{1.5cm}
\begin{abstract} 
We compute the onium-onium scattering amplitude at fixed impact parameter
in the framework of the perturbative QCD dipole model. Relying on conformal
properties of the dipole cascade and of the elementary dipole-dipole 
scattering amplitude, we obtain an exact result for this onium-onium
 scattering amplitude, which is proven to be identical to the BFKL result,
 and which 
exhibits the frame invariance of the calculation. The asymptotic
expression for this amplitude and for the dipole distribution in an onium
at fixed impact parameter agree with previous numerical simulations.
We show how it is possible to describe onium-$e^{\pm}$ deep inelastic 
scattering in the dipole model, relying on $k_T$-factorization properties.
The elementary scattering amplitudes involved in the various processes
are computed using eikonal techniques.

\end{abstract}

\end{titlepage}
\mbox{}
\thispagestyle{empty}
\newpage
\setcounter{equation}{0}
\setcounter{page}{1}
\section{Introduction}
\label{introduction}

The HERA experiments \cite{h1,zeus} have focused attention on the Balitsky-Fadin-
Kuraev-Lipatov (BFKL) Pomeron \cite{lip}, which should be relevant for describing
 the small-$x_{bj}$ behaviour of the proton structure function. More generally,
 this perturbative QCD hard Pomeron describes the behaviour of hadronic scattering
 amplitudes at very high energy $s$ and fixed momentum transfer $t \sim -m^2$ 
($m$ being an hadronic mass scale). In this Leading Logarithmic Approximation,
 one takes into account the exchange of a bound state of two reggeized gluons in
 $t$-channel. This resummation, based on perturbative Regge physics, predicts an 
increase of the amplitude
\begin{equation}
\label{ABFKL}
A_{BFKL} \propto s^{\alpha_P},
\eq
where
\beq
\label{alphap}
\alpha_P = 1 + \frac{\alpha_S N_c}{\pi} 4 \ln 2 > 1.
\eq
Using the optical theorem, it follows that this behaviour violates
 the Froissart bound at very high $s$
\beq
\label{bornefroissart}
\sigma_{tot} \leq c \, \ln^2 s.
\eq
This violation is directly related to the unitarization problem of QCD, which 
is one of the main problems to be solved in the theory of strong interaction.
Various approaches have been recently proposed in order to restore unitarity.
In the multiregge approach, it has been shown that the Generalized Leading 
Logarithmic Approximation \cite{bBJKP, jBJKP, kpBJKP}, where one takes into 
account the exchange of any fixed number of reggeized gluons in $t$-channel, is
 equivalent to the non-compact Heisenberg XXX spin chain \cite{lipxxx, fk} in
 the
 multicolor limit of QCD. The solution of this integrable model is still an 
open 
problem \cite{k, kbkw, wal4, janik1, janik2}.
In the model recently developped by Mueller et 
al \cite{mueller94, muellerpatel,
 muellerunitarite, muellerchen} and separately by Nikolaev et al 
\cite{ nik, nikital}, in order to control the perturbative approach, one deals
 with 
onia, which are heavy quark-antiquark bound states, so that their transverse
 size 
naturally 
provides an infra-red cut-off. The relevant degrees of freedom at high energy
 are
then
made of color dipoles. These color dipoles produce a classical cascade in the 
multicolor limit, which reveals a Pomeron type dynamics. This approach, 
combined with $k_T$-factorization \cite{catani, collins, levin},  has been
 successfully
 applied to deep inelastic $e^\pm-p$ scattering for describing HERA data 
for $F_2$ \cite{npr, nprw, wal7}.  In the more general case of onium-onium
scattering, the BFKL approximation corresponds to the 
exchange of one pair of gluons between two excited dipoles, each one being
extracted
 from one of the two onia.   
The unitarization problem can also be studied in this dipole model. In 
real physics, unitarity and analyticity of the $S$ matrix and the
finite range of strong interaction lead to the Froissart bound 
(\ref{bornefroissart}) for total cross-section. Unitarity implies in
 particular at fixed impact parameter that 
 the probability of any event
 cannot exceed 1, that is 
\beq
|S(b)| \leq 1.
\eq
One way of enforcing unitarity in the framework of perturbative QCD would
then be to take into account two and more left moving dipoles scattering off
 an
 equal number of right moving dipoles. As long as the relative rapidity $Y$ of
 the two onia is not too 
high, it is possible to deal with unitarization effects without facing
 saturations effects, that is one can consider each dipole in the wave function
 of the onia as still dilute \cite{muellerunitarite, kmw}. However, because
 of 
 non-trivial dynamics in
transverse space, the usual Glauber multiple scattering series, which
 corresponds
 to summing up multiple Pomeron exchanges, diverges factorially. A possible 
way of
 escaping this problem is to sum over to the number 
of 
exchanged Pomerons, and only after this summation is performed, averaging over
 configurations of dipoles in the onia wave functions. Such an approach, except
 for 
a toy model where the transverse dynamics is absent \cite{muellerunitarite},
 has not been
 yet carried out analytically. Numerically, such an approach is possible, and
 Monte Carlo simulations indeed
show that QCD unitarizes in the dipole framework \cite{salam, muellersalam}. 

Our aim is to study how this can be carried out analytically. In this paper, we
will focus on the onium-onium cross-section at fixed impact parameter.  
It is organized as follows.
In section {\bf \ref{onium2}}, we calculate the
 onium-onium scattering amplitude at fixed impact parameter. The method
 developped here is rather general, and is based on expansions over conformal
 three points 
correlation functions. Using this representation, we calculate the 
distribution 
of dipoles inside
an onium, at fixed impact parameter, and obtain its precise dependence with
respect to the transverse size of the excited dipole. Then, the 
cross-section at fixed impact parameter is obtained using
the elementary dipole-dipole cross-section, which is calculated by eikonal
 methods and expanded on a conformal basis. Our result
 shows the exact equivalence between the dipole and the BFKL approaches for
 inclusive processes. In section {\bf \ref{e-p}}, we study some physical
 applications of
 the dipole  picture, namely $e^\pm - {\cal O}nium$ deep inelastic scattering.
 We rely on  
$k_T$-factorization and apply eikonal techniques for computing the elementary 
dipole-gluon cross-section.

\section{Onium-onium cross-section at fixed impact parameter}
\label{onium2}
In this section, we calculate the onium-onium scattering amplitude in the BFKL
 approximation. This calculation has been first carried out in
 Ref. \cite{muellerpatel, muellerunitarite}. It used an expansion on a
 conformal
 three 
points correlation function basis. Asymptotic expressions of this basis were
 used
 in this calculation. The calculation which is presently developped  gives the
 correct dependence
with respect to the transverse sizes of onia. This dependence is highly non
 trivial and could not be obtained without a careful treatment of the conformal
 basis.  

We thus consider onium-onium scattering in the leading logarithmic
 approximation.
 The infinite-momentum wave function of an onium, in the large $N_c$ limit, was
 calculated perturbatively in Ref. \cite{mueller94}. This calculation is based
 on an
  approximation of eikonal type, due to the ordering of longitudinal momenta.
 This
 allows one to compute the probability of emitting a gluon from a quark, and
 then
 from a quark-antiquark pair, that is from
a color dipole. In the multicolor limit, the cascade of emitted soft gluons
decouples, leading to a semi-classical cascade of color dipoles, in
term of probability, since interference terms cancels in this limit. The 
onium-onium elastic scattering amplitude at fixed impact parameter $A(Y,b)$
 can then be expressed using a parton type 
formulation. It involves the number of dipoles in each onium and the
 elementary 
cross-section of two such dipoles. For a relative rapidity $Y$ 
and impact parameter $b$, $A(Y,\underline{b})$ can be expressed as
\beq
\label{Ab}
A(Y,\underline{b}) = -i \int d^2 \underline{x}_1
 \int d^2 \underline{x}_2 \int^1_0 dz_1 \int^1_0 dz_2\, 
 \Phi(\underline{x}_1,z_1) \Phi(\underline{x}_2,z_2)
 F(\underline{x}_1,\underline{x}_2,\tilde{Y},\underline{b}).
\eq
$\Phi(\underline{x}_i,z_i)$ is the square of the heavy quark-antiquark part of 
the onium
 wavefunction, $\underline{x}_i$ being the transverse size of the 
quark-antiquark pair
 and $z_i$ the longitudinal momentum fraction of the antiquark. The momentum
$p_1^+$ and $p_2^-$ of the two onia are supposed to be large, with
$\underline{p}_1 = \underline{p}_2 = 0.$ $Y$
is related to $\tilde{Y}$ by $\tilde{Y} = Y + \ln z_1 z_2,$ due to the fact
 that the 
{\it perturbative} dipole cascade originates from the quark-antiquark pair. The 
distribution $\Phi(\underline{x},z)$ of this pair cannot be computed
 perturbatively, and goes far beyond the purpose of the present approach.
    The scattering
amplitude $F$ is then evaluated in term of the perturbative dipole cascade.
Following Ref. \cite{mueller94, muellerpatel},
we define 
$n(x_{01},x,b,\tilde{Y})$ such that 
\beq
\label{defn}
N(\underline{x},\underline{b},Y) = \int d^2 \underline{x} \int^1_0 dz_1
\,  \Phi(\underline{x},z_1) \, n(\underline{x},\underline{x}',\tilde{Y},
\underline{b})
\eq
is the number density of dipoles
 of transverse size $\underline{x}'$,
at a tranverse distance $\underline{b}$ from the center of the quark-antiquark pair, where
 the momentum fraction of the softest of the two gluons (or quark or 
antiquark) which compose the
 dipole
 is larger or equal to $e^{-Y}.$ $\tilde{Y}$ is the relative rapidity with
respect to the heavy quark given by $\tilde{Y} = Y + \ln z_1.$ 
In the leading logarithm
approximation (noted $F^{(1)}$) where the scattering is due to the exchange
 of a single pair of gluons between the two dipoles extracted from the left
 and right moving onia, $F^{(1)}$ reads
\beqa
\label{F1b1}
&& \hspace{-1cm} F^{(1)}(\underline{x}_1,\underline{x}_2,\tilde{Y},
\underline{b}) =
 -\de \int \frac{d^2 \underline{x}'_1}
{2 \pi {x'_1}^2} \frac{d^2 \underline{x}'_2}{2 \pi {x'_2}^2}
 d^2 \underline{b}_1 
\,d^2 \underline{b}_2 \, d^2 (\underline{b}'_2 - \underline{b}'_1) \,\delta^2
(\underline{b}_1 - \underline{b}_2 - \underline{b}'_1 + \underline{b}'_2 - 
\underline{b}) \nonumber \\
&&\times \,  n(\underline{x}_1,\underline{x}'_1,\tilde{Y}_1,\underline{b}_1) \,
 n(\underline{x}_2,\underline{x}'_2,\tilde{Y}_2,\underline{b}_2)
 \, \sigma_{DD}(
\underline{x}'_1,\underline{x}'_2,\underline{b}'_1 - \underline{b}'_2) .
\eqa
The rapidities $\tilde{Y}_1$ and $\tilde{Y}_2$ are such that 
$\tilde{Y}=\tilde{Y}_1 +
 \tilde{Y}_2.$
Eq. (\ref{F1b1}) involves the elementary dipole-dipole cross-section at fixed
 impact
 parameter $\sigma_{DD}$,
 which has been evaluated in \cite{muellerpatel}, and which is
 calculated
 in appendix \ref{calculsigmaddq} using
eikonal techniques.
For two dipoles of transverse sizes $\underline{x}'_1$ and $\underline{x}'_2$,
 whose centers are located at $\underline{b}'_1$ and $\underline{b}'_2,$ one
obtains
\beq
\label{efficaceddrappel}
\sigma_{DD}(
\underline{x}'_1,\underline{x}'_2,\underline{b}'_1 - \underline{b}'_2) = 
\alpha_s \,  \left\{ \ln \frac{|\underline{b}'_1 - \underline{b}'_2 + \frac{
\underline{x}'_1 + \underline{x}'_2}{2}||\underline{b}'_1 - \underline{b}'_2 - 
\frac{\underline{x}'_1 + \underline{x}'_2}{2}| }  {|\underline{b}'_1 - 
\underline
{b}'_2 + \frac{\underline{x}'_1 - \underline{x}'_2}{2}||\underline{b}'_1 - 
\underline{b}'_2 - \frac{\underline{x}'_1 - \underline{x}'_2}{2}| }\right\}^2.
\eq
The forward scattering amplitude can then be evaluated by integration over 
impact parameter, namely
\beq
\label{AAb}
A(Y) = \int d^2\underline{b} \, A(Y,\underline{b}).
\eq 
$A(Y)$ is normalized so that the optical theorem reads
\beq
\label{optique}
\sigma(Y) = 2 \, {\rm Im} A.
\eq
In order to get the expression for $n(\underline{x},\underline{x}',
\tilde{Y},
\underline{b})$, one relies on the
global conformal invariance of the dipole emission kernel, related to the
 absence of
 scale. We thus decompose this distribution on the basis of conformaly
 invariant
 three points
 holomorphic and antiholomorphic correlation functions 
\cite{lipatov86, polyakov}.
 Introducing complex 
coordinates in the two-dimensional transverse space
\beqa
&&\underline{\rho}= (\rho_x,\rho_y)\\
&&\rho = \rho_x + i\rho_y \mbox{ and } \rho^* = \rho_x - i\rho_y,
\eqa
the complete set of eigenfunctions $E^{n,\nu}$ of the dipole emission kernel 
is
\beq
\label{defE}
E^{n, \nu}(\underline{\rho}_{10},\underline{\rho}_{20}) = (-1)^n\left(\frac{
\rho_{12}}{\rho_{10} \rho_{20}} \right)^h \left(\frac{\rho^*_{12}}{\rho^*_{10}
 \rho^*_{20}} \right)^{\bar h},
\eq
with
\beqa
h &=& \frac{1-n}{2} + i \nu \nonumber \\
\bar h &=& \frac{1+n}{2} + i \nu \,
\eqa
being the corresponding conformal weights, with $n$ integer and $\nu$ real.
This set constitutes a unitary irreducible representation of  SL(2,{\bf C})
 \cite{gelfand}.

We get rid of the longitudinal degrees of freedom by using a Mellin transform 
with respect to $\tilde{Y}$, namely
\beq
\label{mellinomega}
n(x,x',\tilde{Y},b) = \int \frac{d \omega}{2 i \pi} e^{\omega \tilde{Y}} n_{\omega}(x,x',b).
\eq
Expanding the dipole distribution on the conformal basis, one writes
\beqa
\label{decn}
&&n_{\omega}(x,x',b) = \sum_{n=-\infty}^{n=+\infty} 8 \int \frac{d \nu}{(2 \pi)^3}
\frac{d^2 w}{x'^2} \left(\nu^2 + \frac{n^2}{4} \right) n_{\{\nu,n\}\omega}
 \nonumber \\
&&\hspace{1cm}\times \,  E^{n,\nu}\left(\bb + \frac{\underline{x}'}{2} - \wb,\bb-
\frac{\underline{x}'}{2} - \wb \right) E^{n,\nu*}\left(\frac{\underline{x}}{2} -
 \wb,-\frac{\underline{x}}{2} - \wb \right).
\eqa
In this expression, the transverse integration is done with respect to the
 coordinate $\underline{w}$ of the center of mass of the quark-antiquark pair. 
The longitudinal dynamics gives rise to the term $n_{\{\nu,n\}\omega},$
which was computed in Ref. \cite{mueller94}, and has the following expression
\beq
\label{distn1}
n_{\{\nu,n\}\omega}= \frac{2}{\omega - \frac{2 \alpha_S N_c}{\pi} \chi(n,\nu)},
\eq
where
\beq
\label{chinnu}
\chi(n,\nu) = \psi(1) - \de \psi\left(\frac{|n|+1}{2} + i \nu\right) -\de \psi
\left(\frac{|n|+1}{2} - i \nu\right) = \psi(1) - {\rm Re} \, 
\psi\left(\frac{|n|+1
}{2} + i \nu\right).
\eq
Going back to the longitudinal space, the distribution of dipoles takes the
 form
\beqa
\label{distn2}
\hspace{-2cm} n(\underline{x},\underline{x}',\tilde{Y},\underline{b}) &=& 
\sum_{n=-\infty}^{+\infty} 16 \int \frac{d \nu}{(2 \pi)^3}
 \frac{d^2 \underline{w}}{x'^2} \left(\nu^2 + \frac{n^2}{4} \right) \exp\left(
\frac{2 \alpha N_c}{\pi} \chi(n,\nu) \tilde{Y}\right) \nonumber \\
&& \hspace{1cm}\times \, E^{n,\nu}\left(\bb + \frac{\xb'}{2} - \wb,\bb-
\frac{\xb'}
{2} - \wb\right) E^{n,\nu*}\left(\frac{\xb}{2} - \wb,-\frac{\xb}{2} - \wb
\right).
\eqa
This distribution can be more easily computed by using a Fourier transform 
with 
respect to the impact parameter, that is by fixing the $t$ channel exchanged
 momentum.
We thus define  
\beq
\label{nq1}
n(\underline{x},\underline{x}',\tilde{Y},\underline{b}) = \int \frac{d^2 \underline{q}}
{(2 \pi)^2} e^{\displaystyle -i \qb.\bb} \, n(\underline{x},\underline{x}',\tilde{Y},
\underline{q}),
\eq
and introduce, following Ref. \cite{lipatov86}, the corresponding mixed 
representation of $E^{n,\nu}$, namely
\beq
\label{Eq1}
E_q^{n,\nu}(\rho) = \frac{2 \pi^2}{b_{n,\nu}} \int \frac{d^2 \underline{R}}
{|\rho|}e^{\displaystyle i \qb.\Rb} E^{n,\nu}\left(\underline{R}+\frac{
\underline{\rho}}{2},\underline{R}-\frac{\underline{\rho}}{2} \right).
\eq
The normalisation term $b_{n,\nu}$ is given by
\beq
\label{bnnu}
b_{n,\nu} = \pi^3 \frac{1}{-i \nu +|n|/2}2^{4 i \nu} \frac{\Gamma(-i \nu +
 (1+|n|)
/2)}{\Gamma(i \nu + (1 + |n|)/2)} \frac{\Gamma(i \nu + |n|/2)}{\Gamma(-i \nu +
|n|/2)}.
\eq
In this Fourier representation the dipole distribution reads
\beq
\label{nq2exact}
n(\underline{x},\underline{x}',\tilde{Y},\underline{b}) =\sum_{n=-\infty}^{+\infty} 
 \int
 \frac{d \nu}{2 \pi} \int \frac{d^2 \underline{q}}{(2 \pi)^2} e^{\displaystyle
 -i
 \underline{q} . \underline{b}} \, E^{n,\nu*}_q(x) E^{n,\nu}_q(x') \frac{x}{x'}
 \exp\left(\frac{2 \alpha N_c}{\pi} \chi(n,\nu) \tilde{Y}\right).
\eq
\subsection{The large rapidity dipole density at fixed impact parameter}
\label{resultatasymptotique}

At very large $\tilde{Y}$, the term corresponding to $n=0$ dominates the exponential
term in the expansion
(\ref{nq2exact}) because of the
expression of $\chi(n,\nu)$ (see Eq. (\ref{chinnu})), and we will restrict
 ourselves to this case in this subsection. 

We first give a useful representation for $E^{0\nu*}_q$ (a general
 representation for $E^{n\nu*}_q$ can also be obtained \cite{np}).
From Eqs. (\ref{defE}) and (\ref{Eq1}), we get
\beqa
\label{Eq2}
E^{0,\nu}_q(\rho)&=& \frac{2 \pi^2}{b_{0,\nu}}\rho^{2 i \nu} \int \frac{d^2
 \underline{R} \, e^{\displaystyle i \underline{q}.\underline{R}}}{\left\{
 \left|\underline{R}-
\underline{\rho}/2\right| \left|\underline{R}+
\underline{\rho}/2\right|\right \}^{1+2 i \nu}} \nonumber \\
&=&  \frac{2 \pi^2}{b_{0,\nu}}\rho^{2 i \nu}\, e^{\displaystyle i \underline{q}.
\underline{\rho}/2} \int \frac{d^2 \underline{R}' \, e^{\displaystyle i 
\underline{q}.\underline{R}'}}{\left\{ \left|\underline{R}'\right| \left|
\underline{R}'+
\underline{\rho}\right|\right \}^{1+2 i \nu}}.
\eqa
Introducing the Feynman representation of the integrand
\beq
\label{DenFeynman}
\frac{1}{\{  \left|\underline{R}'\right|^2 \left|\underline{R}'+
\underline{\rho}\right|^2 \}^{\de+ i \nu}} = \frac{\Gamma(1 + 2 i \nu)}{\Gamma^2
\left(\de + i \nu \right)} \left. \int^1_0 \frac{[\alpha(1-\alpha)]^{i \nu - \de}
 d \alpha}{D_{\alpha}^{1 + 2 i \nu}} \right.,
\eq
where
\beq
\label{Den}
D_{\alpha}= (1-\alpha) \Rb^{'2} + \alpha(\Rb'+\underline{\rho})^2 = |\Rb' +
 \alpha \underline{\rho}|^2 + \underline{\rho}^2 \alpha (1 - \alpha),
\eq
and setting $\Rb = \Rb' + \alpha \underline{\rho},$ Eq. (\ref{Eq2}) now reads
\beq
\label{Eq3}
E^{0,\nu}_q(\rho)= \frac{2 \pi^2}{b_{0,\nu}}\rho^{2 i \nu} \frac{\Gamma(1 + 2 i 
\nu)}{\Gamma^2\left(\de + i \nu \right)} \int^1_0 d \alpha [\alpha(1-\alpha)]^{i
 \nu - \de}  e^{i \underline{q}.\{\underline{\rho}/2 - \alpha \underline{\rho}\}}
 \int \frac{d^2 \underline{R} \, e^{i \underline{q}.\underline{R}}}{\left[
 \underline{R}^2 + \underline {\rho}^2 \alpha(1-\alpha) \right]^{1 + 2 i \nu}}.
\eq 
The integration with respect to the angle $(\underline{q},\underline{R})$ leads
to a Bessel function $J_0(qR)$. The integration with respect to
 $R=|\underline{R}|$ can then be carried out,
using the formula (6.565) of Ref. \cite{Grad}
\beq
\label{Grad}
\int \frac{J_{\nu}(qx) \, x^{\nu+1}}{[x^2 + a^2]^{\mu+1}} d x = \frac{a^{\nu-\mu}
 q^{\mu} K_{\nu-\mu}(aq)}{2^{\mu}\Gamma(\mu+1)}.
\eq
This finally yields
\beq
\label{Eq4}
 E^{0,\nu}_q(\rho)= \frac{4 \pi^3}{b_{0,\nu}} e^{\displaystyle i \underline{q}.
 \underline{\rho}/2} \left(\frac{q}{2}\right)^{2 i \nu} \frac{1}{\Gamma^2\left(
\de + i \nu \right)} \int^1_0 d \alpha [\alpha(1-\alpha)]^{ - \de}  e^{
\displaystyle -i \alpha \underline{q}. \underline{\rho}} K_{-2 i \nu}\left(q 
\rho \sqrt{\alpha(1-\alpha)}\right).
\eq 
This last integration can be performed and yields (see appendix \ref{calculE})
\beq
\label{Eq5}
E^{0,\nu}_q(\rho)= \left(\frac{q}{2}\right)^{2 i \nu} 2^{-2 i \nu}  \Gamma^2(1-i
 \nu) \,
\left[ J_{i \nu}\left(\frac{\rho q}{4} e^{i \Psi}\right) \,
 J_{i \nu}\left(\frac{\rho q}{4} e^{-i \Psi}\right) -  J_{-i \nu}\left(\frac{\rho
 q}{4} e^{i \Psi}\right) \,
 J_{-i \nu}\left(\frac{\rho q}{4} e^{-i \Psi}\right)\right],
\eq
where $\Psi$ is the angle $(\underline{q},\underline{\rho}).$
 We will not need this explicit form in the following calculation.
\\

Let us now compute the contribution to the density of dipole 
(\ref{nq2exact}) corresponding to $n=0$.
Since the longitudinal and the transverse degrees of freedom can be easily 
separated (see formula (\ref{nq2exact})), we will forget the exponential term 
depending on the rapidity $\tilde{Y}$, and will restore this dependence at the very 
end.
We thus define $n_{\{n,\nu\}}(x,x',b)$ such that
\beq
\label{defnnu}
n(\underline {x},\underline {x}',\tilde{Y},\underline {b})= \sum_{n=-\infty}^{+\infty}
 \int \frac{d \nu}{2 \pi} 
n_{\{n,\nu\}}(x,x',b) \exp\left(\frac{2 \alpha N_c}{\pi} \chi(n,\nu) \tilde{Y}\right).
\eq
In the asymptotic case we are interested in,
\beq
\label{defnnuapprox}
n(\underline{x},\underline{x}',\tilde{Y},\underline{b}) \sim  \int \frac{d \nu}{2 \pi} 
n_{\{0,\nu\}}(x,x',b) \exp\left(\frac{2 \alpha N_c}{\pi} \chi(0,\nu) \tilde{Y}\right).
\eq
We can now express the distribution of dipole, taking into account Eq.
 (\ref{defnnuapprox}).
 When $\tilde{Y} \to \infty,$ one can use a saddle approximation for the
 $\nu-$integration. The function $\chi(0,\nu)$ is maximum at $\nu=0,$ 
which thus defines the saddle-point. We will thus expand $n_{\{0,\nu\}}$
around $\nu=0.$ 
Using the expression (\ref{Eq4}), $n_{\{0,\nu\}}$ then reads
\beqa
&& \hspace{-.7cm} n_{\{0,\nu\}}(x,x',b) = \frac{16 \pi^6}{|b_{0,\nu}|^2}
 \frac{x}
{x'} \frac{1}{\Gamma^2\left(\de + i \nu \right)\Gamma^2\left(\de - i \nu 
\right)}
 \int \frac{d^2\underline{q}}{(2 \pi)^2} \int^1_0 d \alpha \int^1_0 d \beta  \,
 [\alpha(1-\alpha)]^{ - \de}  [\beta(1-\beta)]^{ - \de} \nonumber \\
&& \times  \, e^{\displaystyle i \underline{q}. \frac{\underline{x}}{2}(1-2 
\alpha)} e^{\displaystyle -i \underline{q}. \frac{\underline{x}'}{2}
(1-2 \beta)}
 e^{\displaystyle -i \underline{q}.\underline{b}}  K_{2 i \nu}(q x
 \sqrt{\alpha
(1-\alpha)}) K_{-2 i \nu}(q x'\sqrt{\beta(1-\beta)}).
\label{nnu1}
\eqa
Setting $\underline{v} = \frac{\underline{x}}{2}(1-2 \alpha) -
 \frac{\underline{x}'}{2}(1-2 \beta) + \underline{b}$, the integration with
 respect to the angle
$(\underline{q},\underline{v})$ gives a Bessel function $J_0(qv)$.
The integration with respect to $q$ can be carried out using the following 
formula
 \cite{Magnus}
\beqa
\label{mag1}
&&\hspace{-2cm} \int_0^{\infty} x^{\mu+1} K_{\lambda}(a_1 x) K_{\lambda}(a_2 x)
 J_{\mu}(a_3 x) d x \nonumber \\
&& = \frac{1}{2} \sqrt{\frac{\pi}{2}} \left(\frac{a_3}{a_1 a_2}\right)^{\mu+1} 
{\cal P}^{-\mu-\de}_{\lambda-\de} (z) (z^2-1)^{-\frac{\mu}{2} - \frac{1}{4}} 
\Gamma(\mu+\lambda+1) \Gamma (\mu-\lambda+1),
\eqa
where  $\displaystyle z = \frac{1}{2} \frac{a_1^2+a_2^2+a_3^2}{a_1 a_2},$
and ${\cal P}$ is a  Legendre function.
Here  $\mu = 0,$ $\lambda = 2 i \nu$
and the corresponding Legendre function is
\beq
{\cal P}^{-\de}_{\gamma}(z) = (2 \pi)^{-\de} \left(\de+ \gamma\right)^{-1}
 (z^2-1)^{-\frac{1}{4}} \left \{\left [z +(z^2-1)^{\de}\right ]^{\gamma+\de} -
 \left [z +_ (z^2-1)^{\de}\right ]^{\gamma+\de} \right\} \nonumber.
\eq
In the case we are interested in, $a_1=x \sqrt{\alpha(1-\alpha)}$, $a_2= x'
 \sqrt{\beta(1-\beta)}$
and $a_3 = b$.  In the domain $b \gg x,x'$ (where the total cross-section gets 
its dominant contribution: see subsection \ref{resultatexact}), 
\beq
\label{zapprox}
z \simeq \displaystyle \frac{b^2}{2 x x'}[\alpha(1-\alpha)
\beta(1-\beta)]^{-\de} \gg 1,
\eq
 and thus 
\beqa
\label{mag2}
&& \hspace{-.5cm} \int q \,  K_{2 i \nu}\left(q x \sqrt{\alpha(1-\alpha)}\right)
 K_{-2 i \nu}\left(q x'\sqrt{\beta(1-\beta)}\right) J_0(qv)\, d q \simeq \frac{1}
{2}
\frac{1}{2 i \nu}  \Gamma(1+2i \nu) \Gamma(1-2 i \nu)  \nonumber \\
&& \times \frac{1}{b^2} \left\{\left[\frac{b^2}{x x'}\right]^{2 i \nu}
 [\alpha(1-\alpha)\beta(1-\beta)]^{-i\nu}  - \left[\frac{b^2}{x x'}\right]^{-2 i
 \nu}
 [\alpha(1-\alpha)\beta(1-\beta)]^{i\nu}\right\}.
\eqa
Using Ref. \cite{lipatov86}, 
\beq
\label{annu}
a_{n,\nu} = \frac{\pi^4/2}{\nu^2 + n^2/4} = \frac{|b_{n,\nu}|^2}{2 \pi^2},
\eq
which gives for $n_{\{0,\nu\}}$
\beqa
\label{nnu2}
&&\hspace{-1cm} n_{\{0,\nu\}}(\underline{x},\underline{x}',\underline{b}) \simeq 
 -\frac{2 i \nu}{\pi} \frac{x}{x'}
 \frac{1}{b^2} \frac{\Gamma(1+2 i \nu)  \Gamma(1-2 i \nu)}
{\Gamma^2(\de + i \nu)\Gamma^2(\de - i \nu)}\int_0^{\infty} d \alpha \int_0^
{\infty}
 d \beta \nonumber \\
&&\hspace{-.4cm} \times  \left\{\left(\frac{b^2}{x x'}\right)^{2 i \nu} 
 [\alpha(1-\alpha)\beta(1-\beta)]^{-\de -i\nu} 
-  \left(\frac{b^2}{x x'}\right)^{-2 i \nu} [\alpha(1-\alpha)\beta(1-\beta)]^
{-\de
 +i\nu}\right\}.
\eqa
The integration with respect to $\alpha$ and $\beta$ can now be carried out.
 This
 yields
\beq
\label{nnu3}
n_{\{0,\nu\}}(\underline{x},\underline{x}',\underline{b}) \simeq  -\frac{2 i \nu}
{\pi}\frac{x}{x'} \frac{1}{b^2} \left
 \{\frac{\Gamma(1+2 i \nu)}{ \Gamma(1-2 i \nu)}\frac{\Gamma^2(\de - i \nu)}
{\Gamma^2(\de + i \nu)}\left(\frac{b^2}{x x'}\right)^{2 i \nu} - c.c \right\}.
\eq
Using the doubling formula, one finally obtains
\beq
\label{nnu4}
n_{\{0,\nu\}}(\underline{x},\underline{x}',\underline{b}) \simeq  -\frac{2 i \nu}
{\pi}\frac{x}{x'} \frac{1}{b^2} \left
 \{ \left(\frac{16 b^2}{x x'} \right)^{2 i \nu} - \left(\frac{16 b^2}{x x'}
\right)^{-2 i \nu}   + {\cal O}\left(\nu^3\right) \right \}.
\eq
This can now be inserted in Eq. (\ref{defnnuapprox}) to calculate the large
 $\tilde{Y}$ limit of the density, when $b^2/xx' \ll 1.$
 We  develop $\chi(0,\nu)$
 around
 $\nu=0$
\beq
\label{devchi}
\chi(0,\nu) \sim 2 \ln 2 - 7 \zeta(3) \nu^2 + {\cal O}(\nu^4),
\eq
and take into account that $\chi(0,-\nu) = \chi(0,\nu)$.
One has then to compute
\beq
\label{n1}
n(\underline{x},\underline{x}',\tilde{Y},\underline{b}) \simeq  -\frac{x}{x'}
 \frac{1}{b^2} \int^{+\infty}_{-\infty} \frac{d \nu
}{2 \pi} \frac{4 i \nu}{\pi}  \exp\left(\displaystyle \frac{2 \alpha_s N_c}{\pi}2
 \ln 2 \,
 \tilde{Y} - \frac{2 \alpha_s N_c}{\pi} 7 \zeta(3)  \tilde{Y} \,
 \nu^2 + 2 i \nu \ln \frac{16 b^2}{x x'}\right).
\eq
After making the replacement $\nu' = \nu - i \left(\ln \frac{16 b^2}{x x'}
\right)/\left(\frac{2 \alpha_s N_c}{\pi}7 \zeta(3) \tilde{Y}\right)$, the gaussian
integration 
with respect to $\nu'$
 finally yields
\beq
\label{n2}
n(\underline{x},\underline{x}',\tilde{Y},\underline{b}) \simeq \frac{x}{4 b^2 x'} 
\frac{\ln (16 \, b^2/x x')}{\left(\frac{7}{2}
 \alpha_s N_c  \zeta(3) \tilde{Y}\right)^{3/2}} \exp \left \{\displaystyle \frac{4 
\alpha_s N_c}
{\pi}\ln 2 \, \tilde{Y} \right \} \exp \left \{-\displaystyle \frac{\ln^2 (16\, 
 b^2/x x')}
{\frac{14 \alpha_s N_c}{\pi}\zeta(3) \tilde{Y}} \right \}.
\eq
This result is valid in the domain
\beq
\label{domaine}
\frac{2 \alpha_s N_c}{\pi}7 \zeta(3) \tilde{Y} \ll 
\ln \frac{16 b^2}{x x'} \ll 1.
\eq
This expression corrects the expression (8) of Ref. \cite{muellerunitarite}
 when
 one
considers the distribution in transverse space. Numerical simulations and
 approximate
 analytical calculations indeed 
confirm this factor 16 in the transverse distribution \cite{salamthesis}. Note
 that it is only in this asymptotic regime that
 $n(\underline{x},\underline{x}',\tilde{Y},\underline{b})$ has no angular dependence.

\subsection{Exact result for the onium-onium cross-section at fixed impact
 parameter and equivalence with BFKL Pomeron}
\label{resultatexact}
Let us now compute the onium-onium scattering amplitude at fixed impact
 parameter according to the 
 process
 displayed in figure \ref{diffusion}.  We denote $\underline{x}_{a1}$ 
($\underline{x}_{b1}$) the
 transverse
 coordinate of
 the heavy quark (antiquark) making up the right moving onium and
 $\underline{x}_{a2}$ ($\underline{x}_{b2}$) the coordinates of the
 corresponding
quark (antiquark) making the left moving onium. These onia of transverse sizes
 $\underline{x}_1 = \underline{x}_{a1} - \underline{x}_{b1}$ and 
$\underline{x}_2 = \underline{x}_{a2} - \underline{x}_{b2}$ 
 scatter through  the exchange of a pair of
gluons between two elementary dipoles, respectively of transverse size
 $\underline{x}'_1$ and
 $\underline{x}'_2$, located at  $\underline{b}'_1$ and $\underline{b}'_2$
 with respect to the
 reference
 point $\underline{0}$ (which is arbitrary due to translation invariance).
 These two
 elementary dipoles are produced by the two heavy quark-antiquark pairs at a
 distance
 $\underline{b}_1$ and $\underline{b}_2$ from their center of mass. 
\begin{figure}[htb]
\begin{picture}(500,220)(0,0)
\put(-130,130){\epsfysize=6cm{\centerline{\epsfbox{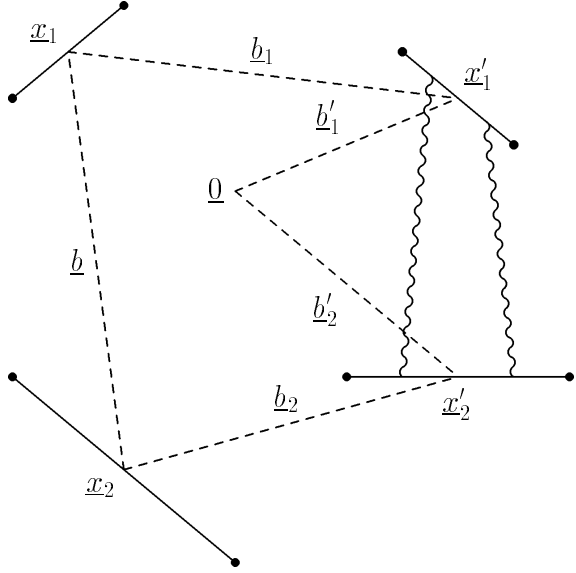}}}}
\end{picture}
\caption{Onium-Onium scattering at leading order.}
\label{diffusion}
\end{figure}

In order to compute $F^{(1)}$ as given by Eq.  (\ref{F1b1}), we now use the
 result (\ref{mellinomega}, \ref{decn}) for the dipole density and the 
following expansion of the dipole-dipole cross-section $\sigma_{DD}$
\beqa
\label{rappelsigmaDD}
&&\sigma_{DD}(\underline{x}'_1,
\underline{x}'_2,\underline{b}'_1-\underline{b}'_2)= 
\frac{2 \alpha_s^2}{(2 \pi)^2} \sum_{n=-\infty}^{+\infty} \int_{-\infty}^{+
\infty}
 d \nu  \int d^2 \underline{w} \left(\nu^2 + \frac{n^2}{4}\right) \frac{1 +
 (-1)^n}{\left( \nu^2 + \left(\frac{n-1}{2} \right)^2 \right) \left( \nu^2 + 
\left ( \frac{n+1}{2} \right)^2 \right)} \nonumber \\
&&\times \, E^{n,\nu*}\left(\underline{b}'_1 + \frac{\underline{x}'_1}{2} - 
\underline{w},\underline{b_1} - \frac{\underline{x}'_1}{2} -
 \underline{w}\right) \,
 E^{n,\nu}\left(\underline{b}'_2 + \frac{\underline{x}'_2}{2} - \underline{w},
\underline{b}'_2 - \frac{\underline{x}'_2}{2} - \underline{w}\right),
\eqa
which is proved in appendix \ref{calculsigmaddq}.
 The full
expression obtained is
\beqa
\label{calculF1}
&&\hspace{-.5cm} F^{(1)}(\underline{x}_1,\underline{x}_2,\tilde{Y},\underline{b})=
 -\frac{\alpha_s^2 (16)^2}{(2 \pi)^2}
 \sum_{n_1 = -\infty}^{+\infty} \sum_{n_2 = -\infty}^{+\infty} \sum_{n =
 -\infty}^{+\infty} \int^{\infty}_{-\infty} \frac{d \nu_1}{(2 \pi)^3}
 \int^{\infty}_{-\infty}
 \frac{d \nu_2}{(2 \pi)^3}  \int^{\infty}_{-\infty} d \nu \int
 \frac{d^2\underline{w}_1}{{x'_1}^2} \int
 \frac{d^2\underline{w}_2}{{x'_2}^2}   \nonumber\\
&&\hspace{-.5cm} \times \, \int d^2 \underline{w} \int \frac{d^2
 \underline{x}'_1}{2 \pi {x'_1}^2} \, \int \frac{d^2 \underline{x}'_2}
{2 \pi {x'_2}^2}\, \int d^2
 \underline{b}_1 \int d^2 \underline{b}_2 \int d^2 \underline{b}_{int}\, 
 \delta^2(\underline{b}_1 - \underline{b}_2 + \underline{b}_{int} -
 \underline{b})
 \,\left(\nu_1^2 + \frac{n_1^2}{4} \right) \, \left(\nu_2^2 + \frac{n_2^2}{4}
 \right) 
 \nonumber \\
&&\hspace{-.5cm}\times \, \left(\nu^2 + \frac{n^2}{4}\right) \,
 \frac{1 + (-1)^n}{\left( \nu^2 + \left(\frac{n-1}{2} \right)^2 \right)  \left(
 \nu^2 +
 \left ( \frac{n+1}{2} \right)^2 \right)} \, \exp
 \left(\frac{2 \alpha_s N_c}{\pi}(\chi(n_1,\nu_1)\tilde{Y}_1
 +\chi(n_2,\nu_2)\tilde{Y}_2)\right) \nonumber \\
&&\hspace{-.5cm}\times \, E^{n_1,\nu_1}\left(\underline{b}_1 +
 \frac{\underline{x}'_1}{2} - \underline{w}_1,\underline{b}_1 -
 \frac{\underline{x}'_1}{2} -
 \underline{w}_1\right) \, E^{n_1,\nu_1*}\left(\frac{\underline{x}_1}{2} -
 \underline{w}_1,-
 \frac{\underline{x}_1}{2} - \underline{w}_1\right) \nonumber \\
&&\hspace{-.5cm}\times \, E^{n_2,\nu_2*}\left(\underline{b}_2 +
\frac{\underline{x}'_2}{2} - \underline{w}_2,\underline{b}_2 -
 \frac{\underline{x}'_2}{2} -
 \underline{w}_2 \right) \,E^{n_2,\nu_2}\left(\frac{\underline{x}_2}{2} -
 \underline{w}_2, -
 \frac{\underline{x}_2}{2} - \underline{w}_2\right) \nonumber \\
&&\hspace{-.5cm}\times \,  E^{n,\nu*}\left(\underline{b}_1 +
 \frac{\underline{x}'_1}{2} - \underline{w},\underline{b}_1 -
 \frac{\underline{x}'_1}{2} - \underline{w}\right)
 \, E^{n,\nu}\left(\underline{b}_{int} + \underline{b}_1 +
\frac{\underline{x}'_2}
{2} -
 \underline{w}_1,\underline{b}_{int} + \underline{b}_1 -
 \frac{\underline{x}'_2}
{2} - 
\underline{w}_1\right) 
\eqa
where we have  used the fact that 
 $\sigma_{DD}(\underline{x}'_1,
\underline{x}'_2,\underline{b}'_1-\underline{b}'_2)$ is translationally
 invariant and only 
depends on $\underline{b}_{int} =\underline{b}'_2 - \underline{b}'_1.$
The quantum numbers $n_1,\nu_1$ and $n_2,\nu_2$
correspond respectively to the dipole distributions
 $n(\underline{x}_1,\underline{x}'_1,
\tilde{Y}_1,\underline{b}_1)$ and $n(\underline{x}_2,\underline{x}'_2,
\tilde{Y}_2,\underline{b}_2)$.

Note the existence of a term $1 + (-1)^n$ which arises from the 
$s \leftrightarrow u$ symetry of the elementary dipole-dipole cross-section
since the gluon is a massless vector boson (see appendix \ref{calculsigmaddq}).
As a consequence only the even $n$ will contribute to the scattering amplitude.

The integration with respect to $\underline{b}_2$ can be performed through the
 delta
 distribution. Then, the remaining expression  can be strongly simplified when
 using the
 following orthonormalization condition for the functions $E^{n,\nu}$ (see Eq.
 (A.16) of Ref.
 \cite{lipatov86})
\beqa
\label{ortho}
&&\hspace{-1cm} \int \frac{d^2 \underline{\rho_1} \,
 d^2 \underline{\rho_2}}{|\rho_{12}|^4} E^{n,\nu}(\rho_{10},\rho_{20}) \,
 E^{m,\mu*}(\rho_{10'},\rho_{20'})= a_{n,\nu}
 \delta_{n,m} \, \delta(\nu -\mu) \, \delta^2(\rho_{00'}) \nonumber \\
&&+ (-1)^n \, b_{n,\nu}|\rho_{00'}|^{-2-4 i \nu} (\rho_{00'}/\rho_{00'}^*)^n
 \delta_{n,-m} \,
 \delta(\nu+\mu).
\eqa
Note that this equation corrects  Eq. (A.16) of Ref.
 \cite{lipatov86} since the factor $(-1)^n$ in the second term of the
 right-hand side was missing. This term arises from the fact that
\beq
\label{signe}
\left(\frac{z^2-1}{z^{2*}-1}\right)^\frac{n}{2} = (-1)^n \left(\frac{1-z^2}
{1-z^{2*}}\right)^\frac{n}{2},
\eq
which has to be taken into account 
when performing the last transformation in Eq. (A.19) of
 Ref.  \cite{lipatov86}.
Applying the relation (\ref{ortho}) 
for $\underline{\rho_1} = \underline{b}_1 + 
\frac{\underline{x}'_1}{2}, \quad  \underline{\rho}_2 = \underline{b}_1 -
 \frac{\underline{x}'_1}{2},
 \quad \underline{\rho}_0 = \underline{w}_1 \mbox{ and } \underline{\rho}_{0'} =
 \underline{w}$ on one hand,
for $\underline{\rho}'_1 = \underline{b}_1 + \underline{b}_{int}
 +\frac{\underline{x}'_2}{2}, \quad  \underline{\rho}'_2 = \underline{b}_1+
\underline{b}_{int} -
 \frac{\underline{x}'_2}{2}, \quad \underline{\rho}_0 = \underline{w}_2 + 
\underline{b}
 \mbox{ and } \underline{\rho}_{0'} = \underline{w}$ on the other hand,
and using the equality  $d^2 \underline{b}_1 \, d^2 \underline{b}_{int} \, d^2
 \underline{x}'_1 \, d^2 \underline{x}'_2 = d^2 \underline{\rho}_1 \,d^2
 \underline{\rho}_2 \,d^2 \underline{\rho}'_1 \,d^2 \underline{\rho}'_2,$ the
 integration
over these transverse variables gives
\beqa
\label{calculF2}
&&\hspace{-.8cm} F^{(1)}(\underline{x}_1,\underline{x}_2,\tilde{Y},\underline{b})=
 -\frac{\alpha_s^2}{(2 \pi)^2}
 \frac{(16)^2}{(2 \pi)^2} \sum_{n_1 = -\infty}^{+\infty} \sum_{n_2 =
 -\infty}^{+\infty} \sum_{n =
 -\infty}^{+\infty} \int^{\infty}_{-\infty} \frac{d \nu_1}{(2 \pi)^3}
 \int^{\infty}_{-\infty}
 \frac{d \nu_2}{(2 \pi)^3}  \int^{\infty}_{-\infty} d \nu \int d^2\underline{w}_1
 \int
 d^2\underline{w}_2  \nonumber\\
&&\hspace{-.8cm} \times \, \int d^2 \underline{w} \,\left(\nu_1^2 + \frac{n_1^2}
{4}
 \right) \, \left(\nu_2^2 + \frac{n_2^2}{4} \right) \, \left(\nu^2 +
 \frac{n^2}{4}\right) \, \frac{1 + (-1)^n}{\left( \nu^2 +
 \left(\frac{n-1}{2} \right)^2 \right) \left( \nu^2 +
 \left ( \frac{n+1}{2} \right)^2 \right)}  \nonumber\\
&&\hspace{-.8cm}\times \, \exp \left(\frac{2 \alpha_s N_c}{\pi}(\chi(n_1,\nu_1)\tilde{Y}_1 
+\chi(n_2,\nu_2)\tilde{Y}_2)\right) \nonumber\\
&&\hspace{-.8cm}\times \,\left[a_{n_1,\nu_1} \delta_{n_1,n} \, \delta(\nu_1 -\nu)
 \,
 \delta^2(\underline{w} -\underline{w}_1 ) + b_{n_1,\nu_1}|\underline{w}
 -\underline{w}_1|^{-2-4 i \nu_1} \left(\frac{w -w_1}{w^* -w_1^*}\right)^{n_1}
 \delta_{n_1,-n} \,
 \delta(\nu_1+\nu) \, (-1)^{n_1}\right ]  \nonumber\\
&&\hspace{-.8cm}\times \, \left[a_{n_2,\nu_2} \delta_{n_2,n} \, \delta(\nu_2 -\nu)
 \,
 \delta^2(\underline{w} -\underline{w}_2 -\underline{b}) + b_{n,\nu}|\underline{w}
 -\underline{w}_2 -\underline{b}|^{-2-4 i \nu}
 \left(\frac{w -w_2 -b}{w^* -w_2^*-b^*}\right)^{n} \delta_{n,-n_2} \right.
 \nonumber\\
&&\hspace{-.8cm}\left. \times  \, \frac{}{}\delta(\nu_2+\nu)^{} \, (-1)^n\right]
 \,
 E^{n_1,\nu_1*}\left(\frac{\underline{x}_1}{2} - \underline{w}_1,-
 \frac{\underline{x}_1}{2} - 
\underline{w}_1\right) \,E^{n_2,\nu_2}\left(\frac{\underline{x}_2}{2} -
 \underline{w}_2, -
 \frac{\underline{x}_2}{2} - \underline{w}_2\right).
\eqa
The two terms involving the normalisation factors $b_{n_1,\nu_1}$ and 
$b_{n,\nu}$
 can
 be reexpressed using the fact that $E^{n,\nu}$ and $E^{n,\nu*}$ are related by
 the 
following relation, which corrects  Eq. (A.12) of Ref. \cite{lipatov86} (see
 appendix \ref{proprieteE})
\beq
\label{lien}
E^{n,\nu*}(\underline{\rho}_{10},\underline{\rho}_{20}) = \frac{b^*_{n,\nu}}
{a_{n,\nu}} \int d^2 \rho_{0'}
 E^{n,\nu}(\underline{\rho}_{10'},\underline{\rho}_{20'}) |\rho_{00'}|^{-2 + 4
 i
 \nu} \left(\frac{\rho^*_{0'0}}
{\rho_{0'0}}\right)^n (-1)^n.
\eq
Note that this relation arises from the equivalence between the two 
corresponding 
representations of $SL(2,C)$ \cite{gelfand}.
The integration with respect to $\underline{w}_1$ and $\underline{w}_2$ can
 then be
 performed. It gives
\beqa
\label{integrew1}
&&\hspace{-1cm}\int d^2 \underline{w}_1 \, E^{n,\nu}\left(\frac{
\underline{x}_1}
{2} -
 \underline{w}_1, - \frac{\underline{x}_1}{2} - \underline{w}_1\right) \,|
\underline{w}
 -\underline{w}_1|^{-2+4 i \nu} \left(\frac{w -w_1}{w^* -w_1^*}\right)^{-n}
 (-1)^{-n}
 \nonumber \\
&&= \frac{a_{n,\nu}}{b^*_{n,\nu}} \, E^{n,\nu*}\left(\frac{\underline{x}_1}{2}
 -
 \underline{w}, - \frac{\underline{x}_1}{2} - \underline{w}\right) .
\eqa
and
\beqa
\label{integrew2}
&&\hspace{-1cm} \int d^2 \underline{w}_2 \, E^{-n,-\nu}\left(
\frac{\underline{x}_2}{2} - \underline{w}_2, - \frac{\underline{x}_2}{2} -
 \underline{w}_2\right) \,|\underline{w} -\underline{w}_2 -\underline{b}|
^{-2-4 i \nu} \left(\frac{w -w_2 -b}{w^* -w_2^*-b^*}\right)^{n} (-1)^n 
\nonumber \\
&&= \frac{a_{-n,-\nu}}{b^*_{-n,-\nu}} \, E^{-n,-\nu*}\left(\frac{
\underline{x}_2}
{2}
 - \underline{w} + \underline{b}, - \frac{\underline{x}_2}{2} - \underline{w} +
 \underline{b}\right),
\eqa
where we have used the fact that $E^{-n,-\nu*}(\rho_{10},\rho_{20})=
 E^{n,\nu}(\rho_{10},\rho_{20}).$
Since $b_{-n,-\nu*} = b_{n,\nu}$ (see Eq. (\ref{bnnu})) and $a_{-n,-\nu} =
 a_{n,
\nu}$
 (see Eq. (\ref{annu})), the contributions of the four terms obtained after
 expanding
 the brackets in Eq. (\ref{calculF2}) are identical and one finally gets, using
 $\tilde{Y}=Y_1 + \tilde{Y}_2,$
\beqa
\label{calculF3}
&&\hspace{-1.4cm} F^{(1)}(x_1,x_2,\tilde{Y},b)= -\frac{\alpha_s^2}{(2 \pi)^2}
{(2 \pi)^6} 
\sum_{n = -\infty}^{+\infty} \int^{\infty}_{-\infty} d \nu   \int d^2 
\underline{w} 
 \left(\nu^2 + \frac{n^2}{4}\right) \frac{1 + (-1)^n}{\left( \nu^2 +
 \left(\frac{n-1}{2} \right)^2 \right) \left( \nu^2 + \left ( \frac{n+1}{2} 
\right)^2
 \right)}  \nonumber\\
&&\hspace{-1.2cm} \times \,  \exp \left(\frac{2 \alpha_s N_c}{\pi}\chi(n,\nu)\tilde{Y}
\right)\,
 E^{n,\nu*}\left(\frac{\underline{x}_1}{2} - \underline{w},- 
\frac{\underline{x}_1}{2}
 - \underline{w}\right) \,E^{n,\nu}\left(\frac{\underline{x}_2}{2} - 
\underline{w}
 +
 \underline{b}, - \frac{\underline{x}_2}{2} - \underline{w} +\underline{b}
 \right).
\eqa

Some comments are in order about this result, obtained without any 
approximation.
First, it is clearly independent of the choice of the reference frame, since
 the
result only depends on the total rapidity $\tilde{Y}.$ In appendix \ref{referentiel} we
explicitly show that this formula describes in an equivalent way, 
 in the laboratory frame of the
left-moving onium (which is defined to be the frame where this onium
is moving relativistically, but has not enough rapidity to reveal its soft
 gluon contents), the scattering
 of
this non evolved  heavy quark-antiquark pair off one excited
dipole at a distance $b$ from the center of mass of the fast right-moving 
 onium. 

Second, and more importantly, 
it explicitly proves the exact equivalence between the dipole and the
 BFKL
approaches at leading order. Indeed, taking into account form factors
when coupling  the $t$-channel bound state of reggeized gluons to the external
 quark-antiquark pairs
 (see appendix \ref{calculsigmaddq} for details) and the difference of 
definition
of amplitudes ($A_{dipole} = \frac{1}{2s} A_{BFKL}$), we
 should have 
the following relation between the dipole and the BFKL result
\beq
\label{equivdipoleBFKL}
F^{(1)}(\underline{x}_1,\underline{x}_2,\tilde{Y},\underline{b}) =  \alpha_s^2 \int
 \frac{d \omega}{2 \pi i} \exp (\omega \tilde{Y}) \,
 [f_{\omega}(\underline{x}_{a1},\underline{x}_{b1},\underline{x}_{a2},
\underline{x}_{b2}) + f_{\omega}(\underline{x}_{b1},\underline{x}_{a1},
\underline{x}_{a2},\underline{x}_{b2})],
\eq
where $f_{\omega}(\underline{x}_{a1},\underline{x}_{b1},\underline{x}_{a2},
\underline{x}_{b2})$ is defined by equation (26) of
 Ref. \cite{lipatov86}
\beqa
\label{fomega}
&& f_{\omega}(\underline{x}_{a1},\underline{x}_{b1},\underline{x}_{a2},
\underline{x}_{b2})= \sum_{n=-\infty}^{+\infty} \int_{-\infty}^{+\infty}
 \int d^2\underline{x}_0 
 \left(\nu^2 + \frac{n^2}{4}\right) \frac{1}{\left( \nu^2 +
 \left(\frac{n-1}{2} \right)^2 \right) \left( \nu^2 + \left ( \frac{n+1}{2} 
\right)^2
 \right)} \nonumber\\
&&\times \, \frac{1}{\omega - \frac{2 \alpha_s N_c}{\pi}\chi(n,\nu)} 
 E^{n,\nu*}\left(\underline{x}_{a2} - \underline{x}_0,
 \underline{x}_{b2} - \underline{x}_0\right)
 \,E^{n,\nu}\left(\underline{x}_{a1} - \underline{x}_0,
 \underline{x}_{b1} - \underline{x}_0 \right).
\eqa
In this formula the integrand gets a factor $(-1)^n$ when permuting
$x_{a1} \leftrightarrow x_{b1}.$ Performing the changes of variable 
$\underline{w} = \underline{x}_0 -
 \frac{\underline{x}_{a1} + \underline{x}_{b1}}{2}$ and
 $(n,\nu) \to (-n,-\nu),$ one then recovers exactly
the expansion (\ref{calculF3}) with the factor $1 + (-1)^n,$ which proves
the result.
Note that the equivalence
 between the BFKL and dipole kernel can be also proven by comparing the real 
and virtual graphs in covariant and light-cone quantization. The result is that
the sum of real and virtual contributions is identical in both case, although
each of these terms differs. Thus, this result is true only for
 inclusive quantities \cite{muellerchen}.
\\

Defining $F^{(1)}_{\{n,\nu\}}$ as
\beq
\label{defFnnu}
F^{(1)}(\underline{x}_1,\underline{x}_2,\tilde{Y},\underline{b}) =
 \sum_{n=-\infty}^{+\infty} \int_{-\infty}^{+\infty} \frac{d \nu}{2 \pi}
 \, F^{(1)}_{\{n,\nu\}}(\underline{x}_1,\underline{x}_2,\underline{b}) \,
 \exp \left(\frac{2 \alpha_s N_c}{\pi} \chi(n,\nu) \tilde{Y} \right),
\eq
and using Eqs. (\ref{distn2}) and (\ref{defnnu}), equation (\ref{calculF3})
 can be rewritten as
\beq
\label{relationFn}
F^{(1)}_{\{n,\nu\}}(\underline{x}_1,\underline{x}_2,\underline{b})= 
-\frac{\pi \alpha_s^2 x_2^2}{8}  \frac{1 + (-1)^n}{\left( \nu^2 +
 \left(\frac{n-1}{2} \right)^2 \right) \left( \nu^2 + \left ( \frac{n+1}{2}
 \right)^2
 \right)} \, n_{\{n,\nu\}}(\underline{x}_1,\underline{x}_2,\underline{b}).
\eq

In the asymptotic regime where one can keep only the term corresponding to 
$n=0,$ this relation simplifies to
\beq
\label{relationFn0}
F^{(1)}_{\{0,\nu\}}(\underline{x}_1,\underline{x}_2,\underline{b})= 
-\frac{\pi \alpha_s^2 x_2^2}{4} \frac{1}{\left( \nu^2 + \frac{1}{4} \right)^2}
 \, n_{\{0,\nu\}}(\underline{x}_1,\underline{x}_2,\underline{b}).
\eq
Computing $F^{(1)}(\underline{x}_1,\underline{x}_2,\tilde{Y},\underline{b})$
by a saddle point method as we did for
 $n(\underline{x}_1,\underline{x}_2,\tilde{Y},\underline{b})$ (see Eq. (\ref{n1})),
one has to expand the prefactor around $\nu =0,$ which yields
\beqa
\label{resFexact}
F^{(1)}(\underline{x}_1,\underline{x}_2,\tilde{Y},\underline{b})&\simeq&
-4 \pi \alpha_s^2
x_2^2 \, n(\underline{x}_1,\underline{x}_2,\tilde{Y},\underline{b}) \nonumber \\
&\simeq&   - \pi \alpha_s^2\frac{x_1 \, x_2}{b^2}
 \frac{\ln (16 \, b^2/x_1 x_2)}{\left(\frac{7}{2} \alpha_s N_c  \zeta(3)
 \tilde{Y}\right)^{3/2}} \exp \left \{\displaystyle
 \frac{4 \alpha_s N_c}{\pi}\ln 2 \, \tilde{Y} \right \} \exp
 \left \{-\displaystyle \frac{\ln^2 (16\,  b^2/x_1 x_2)}{\frac{14 \alpha_s N_c}
{\pi}
\zeta(3) \tilde{Y}} \right \} \nonumber \\
\eqa
in the domain 
\beq
\label{domaineF}
\frac{2 \alpha_s N_c}{\pi}7 \zeta(3) \tilde{Y} \ll 
\ln \frac{16 b^2}{x_1 x_2} \ll 1.
\eq
This result, which differs from Eq. (10) of Ref. 
\cite{muellerunitarite} by a factor $16$, is 
in aggreement with numerical simulations \cite{salamthesis}.

\subsection{Calculation of the onium-onium total cross-section}
\label{efficacetotale}

In this subsection we compute the onium-onium total cross-section.
It is related to the onium-onium  cross-section at fixed impact parameter
by 
\beq
\label{FbFtotal}
F^{(1)}_{tot}(\underline{x}_1,\underline{x}_2,\tilde{Y}) = \int d^2 \underline{b} 
\, F^{(1)}
(\underline{x}_1,\underline{x}_2,\tilde{Y},\underline{b}).
\eq
We show that, provided the elementary dipole-dipole
 cross section is integrated over distances but 
not averaged over angles, one can get the 
onium-onium total cross-section by a simple direct calculation.

Combining formulae (\ref{F1b1}) and (\ref{FbFtotal}), one obtains 
\beqa
\label{Ftotal1}
F^{(1)}_{tot}(\underline{x}_1,\underline{x}_2,\tilde{Y}) &=&-\de \int \frac{d^2 
\underline{x}'_1}{2 \pi {x'_1}^2} \frac{d^2 \underline{x}'_2}{2 \pi {x'_2}^2}
 d^2
 \underline{b}_1 \,d^2 \underline{b}_2 \, d^2 (\underline{b}_2'- 
\underline{b}_1')
  \nonumber \\
&&\hspace{1cm} \times \,  n(\underline{x}_1,\underline{x}'_1,\underline{b}_1,
\tilde{Y}_1)
 \, n(\underline{x}_2,\underline{x}'_2,\underline{b}_2,\tilde{Y}_2) \, \sigma_{DD}(
\underline{x}'_1,\underline{x}'_2,\underline{b}'_1-\underline{b}'_2)
 \nonumber \\
&=& -\de \int \frac{d^2 \underline{x}'_1}{2 \pi {x'_1}^2} \frac{d^2
 \underline{x}'_2}{2 \pi {x'_2}^2}  \,   n(\underline{x}_1,\underline{x}'_1,
\tilde{Y}_1)
 \, n(\underline{x}_2,\underline{x}'_2,\tilde{Y}_2) \int d^2 (\underline{b}_2'- 
\underline{b}_1')\, \sigma_{DD}(
\underline{x}'_1,\underline{x}'_2,\underline{b}'_1-\underline{b}'_2),
\nonumber \\
\eqa
where $n(\underline{x}_i,\underline{x}_i',\tilde{Y}_i)$ is the integrated
 density of dipoles
\beq
\label{nbntotal}
n(\underline{x},\underline{x}',\tilde{Y}) = \int d^2 \underline{b} \, \, n(
\underline{x},
\underline{x}',\tilde{Y},\underline{b}).
\eq
Thus, since $\underline{b}$ and $\underline{q}$ are Fourier conjugated,
\beqa
\label{ntotal1}
n(\underline{x},\underline{x}',\tilde{Y})&=& \sum_{n=-\infty}^{+\infty}  \int \frac{d
 \nu}
{2 \pi} \lim_{q \to 0} E^{n,\nu*}_q(x) E^{n,\nu}_q(x') \frac{|x|}{|x'|} \exp\left(
\frac{2 \alpha N_c}{\pi} \chi(n,\nu) \tilde{Y}\right) \nonumber\\
&=& \sum_{n=-\infty}^{+\infty} \int \frac{d \nu}{2 \pi} \frac{|x|}{|x'|} \left(\frac{x^* x'}{x {x'}^*}
\right)^{n/2} \left|\frac{x'}{x}\right|^{-2 i \nu} \exp\left(\frac{2 \alpha
 N_c}
{\pi} \chi(n,\nu) \tilde{Y}\right) 
\eqa
where we have used Eq. (\ref{nq2exact}) and  the expansion (\ref{devEq}) of
$E_q^{n,\nu}$ for $q \to 0.$
The integration of $\sigma_{DD}(\underline{x}'_1,
\underline{x}'_2,\underline{b}'_1-\underline{b}'_2)$
 with respect to the distance $\underline{b}_2'- 
\underline{b}_1'$ is performed in appendix \ref{calculsigmaddq} and is given by
 formula
 (\ref{sigmaangle}), which still depends on the orientation of the elementary
 dipoles. One now gets for $F^{(1)}_{tot}$
\beqa
\label{Ftotal2}
&&\hspace{-1cm} F^{(1)}_{tot}(\underline{x}_1,\underline{x}_2,\tilde{Y}) = -\de \sum_{n=-
\infty}^{+\infty}
 \int_{-\infty}^{+\infty} d \nu \, \sum_{n_1=-\infty}^{+\infty} \int_{-\infty}
^{+\infty} \frac{d \nu_1}{2 \pi} \, \sum_{n_2=-\infty}^{+\infty} \int_{-\infty
}^
{+\infty} \frac{d \nu_2}{2 \pi} \,  \int \frac{d^2 \underline{x}'_1}{2 \pi 
{x'_1}
^2} \frac{d^2 \underline{x}'_2}{2 \pi {x'_2}^2} \nonumber \\
&&\hspace{-.7cm} \times \, \frac{x_1}{x'_1} \left(\frac{x_1^* x_1'}
{x_1 {x_1'}^*}
\right)^{n_1/2} \left|\frac{x_1'}{x_1}\right|^{-2 i \nu_1} \,
\frac{x_2}{x'_2} \left(\frac{x_2^* x_2'}{x_2 {x_2'}^*}\right)^{n_2/2} \left|
\frac{x_2'}{x_2}\right|^{-2 i \nu_2} \exp\left(\frac{2 \alpha N_c}{\pi} 
(\chi(n_1,
\nu_1) \tilde{Y}_1+ \chi(n_2,\nu_2) \tilde{Y}_2) \right) \nonumber \\
&& \hspace{-.7cm} \times \,  \alpha_s^2 \, \frac{x'_1 x'_2}{4} \,  \frac{1 +
 (-1)^n}{\left( \nu^2 + \left(\frac{n-1}{2} \right)^2 \right) \left( \nu^2 + 
\left ( \frac{n-1}{2} \right)^2 \right)}
\left(\frac{x'_1 {x'_2}^*}{{x'_1}^* x'_2}\right)^{-n/2} 
\left|\frac{x'_1}{x'_2}\right|^{2 i \nu}.
\eqa
Integrating with respect to $x'_1$ and $x'_2$ leads to a lot of $\delta$
 distributions since
\beq
\label{integredelta}
\frac{1}{2 \pi} \int \frac{d x'_1 \, d {x'_1}^*}{2 x'_1 {x'_1}^*} {x'_1}^
{\frac{n_1-n}{2}-i(\nu_1-\nu)} \, {x'_1}^{*- \frac{n_1 - n}{2} -i(\nu_1-\nu)} =
\pi \, \delta_{n_1,n} \, \delta(\nu_1-\nu)
\eq
and a similar integral over $x'_2.$
Defining  $F^{(1)}_{tot\{n,\nu\}}
(\underline{x}_1,\underline{x}_2)$ and 
$n_{\{n,\nu\}}(\underline{x}_1,\underline{x}_2)$ via
\beq
\label{defFquant}
F^{(1)}_{tot\{n,\nu\}}(\underline{x}_1,\underline{x}_2)
= \sum_{n=-\infty}^{+\infty} \int_{-\infty}^{+\infty} \frac{d \nu}{2 \pi}
 \, F^{(1)}_{tot\{n,\nu\}}(\underline{x}_1,\underline{x}_2) \,
 \exp \left(\frac{2 \alpha_s N_c}{\pi} \chi(n,\nu) \tilde{Y} \right)
\eq
and 
\beq
\label{defnquant}
n_{\{n,\nu\}}(\underline{x}_1,\underline{x}_2)
= \sum_{n=-\infty}^{+\infty} \int_{-\infty}^{+\infty} \frac{d \nu}{2 \pi}
 \, n_{\{n,\nu\}}(\underline{x}_1,\underline{x}_2) \,
 \exp \left(\frac{2 \alpha_s N_c}{\pi} \chi(n,\nu) \tilde{Y} \right),
\eq
this finally yields
\beqa
\label{Ftotal3}
F^{(1)}_{tot\{n,\nu\}}(\underline{x}_1,\underline{x}_2) &=& -\frac{\pi \,
 \alpha_s^2}{4} \,   \frac{1 + (-1)^n}{\left( \nu^2 + \left(\frac{n-1}{2}
 \right)^2
 \right) \left( \nu^2 + \left ( \frac{n+1}{2} \right)^2 \right)}   \left(
\frac{x_1^* x_2}{x_1 {x_2}^*}\right)^{n/2} |x_1|^{1 +2 i \nu} \,
|x_2|^{1 -2 i 
\nu}\nonumber \\
&=&  -\frac{\pi \alpha_s^2 \, x_2^2}{8} \,   \frac{1 + (-1)^n}{\left( \nu^2 + \left(
\frac{n-1}{2} \right)^2 \right) \left( \nu^2 + \left ( \frac{n+1}{2} \right)^2 
\right)}  n_{\{n,\nu\}}(\underline{x}_1,\underline{x}_2)
\eqa
Comparison with Eq. (\ref{relationFn}) provides a check of this calculation 
when integrating both sides with respect to $\underline{b}.$
In the asymptotic regime, corresponding to $n=0,$
\beq
\label{relationFntotal0}
F^{(1)}_{\{0,\nu\}}(\underline{x}_1,\underline{x}_2)= 
-\frac{\pi \alpha_s^2 x_2^2}{4} \frac{1}{\left( \nu^2 + \frac{1}{4} \right)^2}
 \, n_{\{0,\nu\}}(\underline{x}_1,\underline{x}_2),
\eq
which could be obtained from Eq. (\ref{relationFn0}).
Using Eq. (\ref{ntotal1}), the corresponding asymptotic integrated dipole 
distribution reads
\beq
\label{ntotalasym1}
n(\underline{x}_1,\underline{x}_2,\tilde{Y})=
\int \frac{d \nu}{2 \pi} \frac{x_1}{x_2}  \left(\frac{x_1}{x_2}\right)^{-2 i
 \nu}
 \exp\left(\frac{2 \alpha N_c}{\pi} \chi(n,\nu) \tilde{Y}\right), 
\eq
which gives, in the saddle point approximation at large $\tilde{Y},$
\beq
\label{ntotalasym2}
n(\underline{x}_1,\underline{x}_2,\tilde{Y})= \frac{1}{2} \frac{x_1}{x_2} 
 \frac{\exp \left \{\displaystyle
 \frac{4 \alpha_s N_c}{\pi}\ln 2 \, \tilde{Y} \right \}}{\sqrt{\frac{7}{2} \alpha_s
 N_c 
 \zeta(3)
 \tilde{Y}}}  \exp
 \left \{-\displaystyle \frac{\ln^2 (x_1/ x_2)}{\frac{14 \alpha_s N_c}{\pi}
\zeta(3) \tilde{Y}} \right \} \,.
\eq
Thus, 
one gets for the total cross-section
\beqa
\label{Ftotal4}
F^{(1)}(\underline{x}_1,\underline{x}_2,\tilde{Y})&\simeq& -4 \pi \alpha_s^2
x_2^2 \, n(\underline{x}_1,\underline{x}_2,\tilde{Y}) \nonumber \\
&\simeq&-2 \pi \, \alpha_s^2 \, x_1 \,x_2 
 \frac{\exp \left \{\displaystyle
 \frac{4 \alpha_s N_c}{\pi}\ln 2 \, \tilde{Y} \right \}}{\sqrt{\frac{7}{2} \alpha_s
 N_c 
 \zeta(3)
 \tilde{Y}}}  \exp
 \left \{-\displaystyle \frac{\ln^2 (x_1/ x_2)}{\frac{14 \alpha_s N_c}{\pi}
\zeta(3) \tilde{Y}} \right \} \,,
\eqa
in agreement with formula (26) of Ref. \cite{muellerpatel}.

Let us integrate the scattering amplitude $F^{(1)}(\underline{b})$ 
with respect to the impact parameter $\underline{b}$ in the domain 
(\ref{domaineF}) where formula (\ref{resFexact}) is valid (neglecting the 
fact that the upper bound is not infinite).
The corresponding integration then gives
\beq
\label{integreFbexact}
F^{(1)}(\underline{x}_1,\underline{x}_2,\tilde{Y})
\simeq -2 \pi \, \alpha_s^2 \, x_1 \,x_2 
 \frac{\exp \left \{\displaystyle
 \frac{4 \alpha_s N_c}{\pi}\ln 2 \, \tilde{Y} \right \}}{\sqrt{\frac{7}{2} \alpha_s 
N_c 
 \zeta(3)
 \tilde{Y}}}  \exp
 \left \{-\displaystyle \frac{\ln^2 (x_1/ x_2)}{\frac{14 \alpha_s N_c}{\pi}
\zeta(3) \tilde{Y}} \right \} \,,
\eq
which is identical to Eq. (\ref{Ftotal4}). From the gaussian distribution
obtained in Eq. (\ref{integreFbexact}), it is clear, comparing with 
Eq. (\ref{resFexact}), that the total cross 
section at BFKL order is dominated by impact parameter configuration 
much larger than the transverse sizes of the two scattering onia,
corresponding to
\beq
\label{ordreb}
\ln \left(\frac{16 b^2}{x_1 x_2}\right) \sim \sqrt {\frac{14 \alpha_s N_c}{\pi}
\zeta(3) \tilde{Y}}.
\eq
Note that this dominant contribution is inside the domain (\ref{domaineF}).
These dominant configurations are  much more central than what was claimed in
 Ref. \cite{muellerunitarite}.  
It confirms previous numerical simulations
 \cite{salam}. Thus, the calculation, based on perturbative 
QCD, is expected to remain valid for high values of $\tilde{Y}.$

\section{Electron-Onium Deep Inelastic Scattering}
\label{e-p}

In this section, we perform an analysis of $e^\pm - onium$ deep inelastic
 scattering
 at low $x_{bj}$,
based on $k_T$-factorization and dipole color model, as illustrated in figure
 \ref{graphef2}.
\begin{figure}[htb]
\centering
\epsfysize=12.0cm{\centerline{\epsfbox{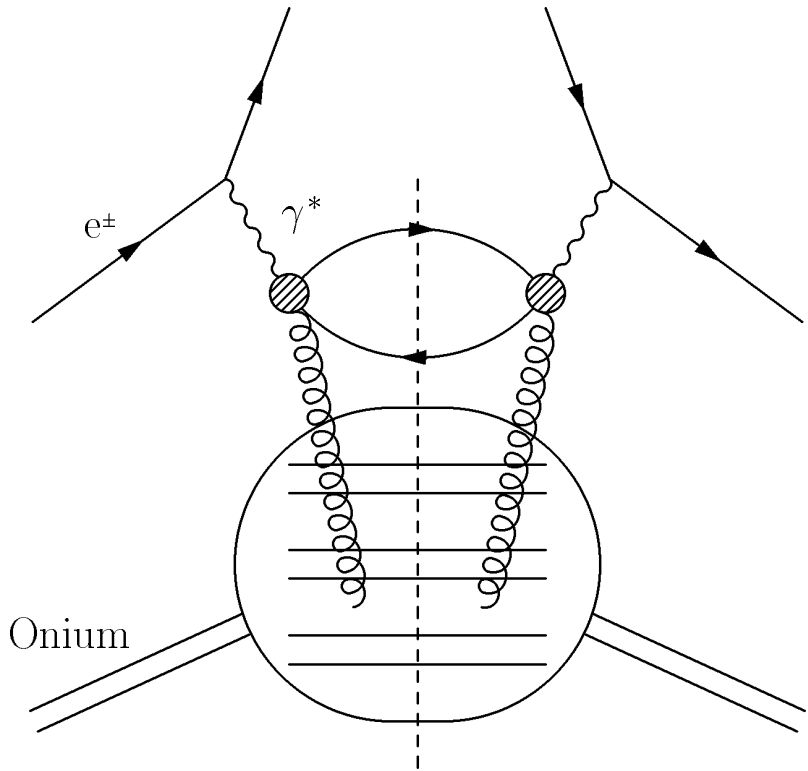}}}
\caption{$k_T$-factorization and dipole model applied to $e^{\pm}-onium$ deep
 inelastic scattering.}
\label{graphef2}
\end{figure}
Our aim is to compute various structure functions in the small $x_{bj}$ 
regime, where the
onium wave function is dominated by a color dipole cascade.

\noindent
In the Regge limit, one can apply the $k_T$-factorization tool
 \cite{catani, collins, levin} in
 order to
extract a photon of virtuality $Q^2$ off an onium. It involves the elementary 
Born  cross-section $\hat{\sigma}_{\gamma g}/Q^2$ of the process  $\gamma~ g(k)
 \rightarrow q ~ \bar{q}$. Here
the gluon is off-shell, quasi transverse, with a virtuality 
$k^2 \, \simeq \kb^2$. One also has to introduce the unintegrated gluon
distribution density at a factorization scale $Q_0^2$, which is related to the
 usual
 gluon distribution by
\beq
\label{gluondist}
G(x_{bj},Q^2,Q_0^2) = \int^{Q^2}_{0} d^2 \kb \, {\cal F}(x_{bj},\kb,Q_0^2).
\eq
We first deal with a dipole of transverse size $x_{01}$, which can be either
part of
 a
 heavy
 onium (i.e. a heavy $q \bar q$ pair) or extracted from a proton as will be
 emphazied
 later. The $k_T$-factorization implies, for the total $\gamma^* - dipole$
 cross-
section $\sigma_{\gamma^*}^{d}$,
\beq
\label{kt1}
Q^2 \sigma_{\gamma^*}^{d} (x_{bj},Q^2;x_{01}^2) = \int d^2\kb \int^{1}_{0}
\frac{d z}{z}\,  \hat{\sigma}_{\gamma g} (x_{bj}/z,\underline k^2/Q^2)
\,
 {\cal F}(z,\kb;x_{01}^2),
\eq
${\cal F}(z,\kb;x_{01}^2)$ being the Fourier transform of 
${\cal F}(z,\kb;Q_0^2)$ in transverse space.
We next evaluate ${\cal F}$ by coupling
 the gluon to the softest dipole which arises in the
 cascade.
This is achieved by using a second $k_T$-factorization.
It involves the elementary
Born  cross-section $\hat{\sigma}_{g d}/k^2$ of the process  $d(\xb)~ g(k)
 \rightarrow  ~ d(\xb)$ for a dipole of transverse size $\xb$ and a soft gluon
 of
 virtuality
 $\kb^2$.
This $k_T$-factorization can be expressed by
\beq
\label{kt2}
\kb^2 {\cal F}(z,\kb;x_{01}^2) = \int \frac{d^2\xb}{(2 \pi)^2\xb^2} 
\int^{z_1}_{z}
\frac{d z'}{z'} \, n (\xb_{01},\xb,\ln \frac{z_1}{z'}) \, \hat{\sigma}_{g d} 
(z/z',\xb^2 \kb^2) \, \delta(z/z'-1),
\eq
where the distribution density $n (\xb_{01},\xb,z)$ 
was defined in section \ref{onium2}.
As previously $z_1 p_+$ is the light-cone
 momentum of
 the quark part of the dipole $(\xb_{01})$.  $\hat{\sigma}_{g d}$ is computed
 in
 appendix \ref{calculsigmagd} using eikonal techniques. Defining 
$\tilde{Y}= \ln z_1/z,$ one gets for the
 dipole-photon cross-section 
\beqa
\label{kt1kt2}
Q^2 \sigma_{\gamma^*}^{d} (x_{bj},Q^2;x_{01}^2) = \int d^2\kb
 \int^{1}_{0}
\frac{d z}{z} \, \hat{\sigma}_{\gamma g} (x_{bj}/z,\frac{\kb^2}{Q^2}) \int
 \frac{d^2\xb}{\xb^2} \, n (\xb_{01},\xb,\tilde{Y}) \nonumber \\
\times 4 \, \pi \, \alpha_s \, \frac{N_c}{(2 \pi)^4} \, (2 - e^{i \kb . \xb} - 
 e^{-i \kb . \xb}) \frac{1}{\kb^2}.
\eqa
Let us now compute the convolutions in longitudinal and transverse spaces.
As in section \ref{onium2}, we introduce a double Mellin-transform in these 
both
variables, namely
\beq
\label{mellinY}
n(\xb_{01},\xb,\tilde{Y}) = \int \frac{d \omega}{2 i \pi} e^{\omega \tilde{Y}}
 n_{\omega}(\xb_{01},\xb)
\eq
and
\beq
\label{mellinxb}
n_{\omega}(\xb_{01},\xb) = \int \frac{d \gamma}{2 i \pi}
 \left(\frac{x_{01}}{x} \right)^{2 \gamma} n_{\omega}(\gamma).
\eq
Here the Mellin variable in transverse space is $\gamma = \frac{1}{2}+i \nu,$
which is introduced here rather than $\nu$ since it plays the role of an
 anomalous
 dimension. We consider only the dominant Regge trajectory, that is $n=0.$
Formula (\ref{distn1}) then reads
\beq
\label{nnuomega}
n_{\omega}(\gamma) = \frac{2}{\omega - \frac{2 \alpha_s N_c}{\pi} \chi(\gamma)}
\eq
where
\beq
\label{chi}
\chi(\gamma) = \chi\left(0,\frac{1}{2} + i\nu\right)=\Psi(1) - \de 
\Psi(\gamma) - 
 \de \Psi(1 - \gamma)
\eq
The quantity
$\hat{\sigma}_{\gamma g}$ has been calculated for different polarizations of 
the
 incoming photon in \cite{catani}.
We introduce the corresponding double Mellin transform of this cross-section,
 namely
\beq
\label{hw}
4 \pi^2 \alpha_{e.m} h_{\omega}(\gamma) = \gamma \int^{\infty}_{0}
 \frac{d \kb^2}{\kb^2} \left(\frac{\kb^2}{Q^2}\right)^{\gamma} 
\hat{\sigma}_{\omega}\left(\frac{\kb^2}{Q^2}\right)
\eq
or equivalently  
\beq 
\label{hinv}
\hat{\sigma}_{\omega}\left(\frac{l^2}{Q^2}\right)= 4 \pi^2 \alpha_{e.m}
 \int^{ }_{ } \frac{d \gamma}{2 i \pi}
 \left(\frac{l^2}{Q^2}\right)^{-\gamma} \frac{1}{\gamma} h_{\omega}(\gamma),
\eq
with
\beq
\label{defsigmaomega}
\hat{\sigma}_{\omega}\left(\frac{\underline{k}^2}{Q^2} \right) = \int^1_0
dz \,z^{\omega-1} \, \hat{\sigma}\left(z,\frac{\underline{k}^2}{Q^2}\right).
\eq
The expression for $\sigma_{\gamma^*}^{d}$ now reads, after performing the
convolution in longitudinal space,
\beqa
\label{kt3}
&&\hspace{-1.5cm} Q^2 \sigma_{\gamma^*}^{d} (x_{bj},Q^2;x_{01}^2) = 
4 \pi^2 \alpha_{e.m} \frac{\alpha_s N_c}{4 \pi^3} \int  d^2\kb 
\int
 \frac{d \gamma'}{2 i \pi}  \int \frac{d \gamma}{2 i \pi} \int
 \frac{d \omega}{2 i \pi} \exp \left( \omega \, \ln \frac{z_1}{x_{bj}} \right)
 \left(\frac{x_{01}}{x}\right)^{2 \gamma} \nonumber \\
&&\times \frac{2}{\omega - \frac{2 \alpha_s N_c}{\pi} \chi(\gamma)} 
 \frac{h_{\omega}(\gamma')}{\gamma'}  \int^{ }_{ } \frac{d^2\xb}{\xb^2} 
 \left(\frac{\underline k^2}{Q^2}\right)^{-\gamma'} (2 - e^{i \kb . \xb} -
  e^{-i \kb . \xb}) \frac{1}{\kb^2} .
\eqa
The integration with respect to the polar angle of $\xb$ 
leads to a Bessel function. One has then to integrate over $x$, namely
\beq
\label{integrationx}
\int \frac{d x}{x} 4 \pi (1 - J_0(kx))\left (\frac{x}{x_{01}} \right)^{-2
 \gamma} =
4 \pi (k x_{01})^{2 \gamma} \frac{2 ^{-1 -2 \gamma}}{\gamma}
 \frac{\Gamma(1- \gamma)}{\Gamma(1 + \gamma)} \equiv 4 \pi (k x_{01})^{2
 \gamma} 
v(\gamma).
\eq
The integration over $k$ gives $\gamma = \gamma'$.
Since $n_{\omega}(\gamma)$ (formula (\ref{nnuomega})) exhibits a pole at 
$\omega_p = \frac{\alpha_s N_c}{\pi} \chi(\gamma)$, the $\omega$ integral 
finally
 yields
\beq
\label{gonium1}
\frac{Q^2} {4 \pi^2 \alpha_{e.m}} \sigma_{\gamma^*}^{d} (x_{bj},Q^2;x_{01}^2)
 = 
\frac{2 \alpha_s N_c}{\pi} \int \frac{d \gamma}{2 i \pi}  
 h_{\omega_p}(\gamma) \frac{v(\gamma)}{\gamma} (\xb_{01}^2Q^2)^{\gamma} 
\exp{\left(\frac{2\alpha_s N_c}{\pi}
 \chi(\gamma) \ln \frac{z_1}{x_{bj}}\right)}.
\eq
In the regime we are interested in, $\omega_p \ll \gamma$ and the dependence 
of $h_{\omega_p}(\gamma)$ on $\omega_p$ can be neglected, replacing $\omega_p$
by 0
 \cite{catani}.
We then get for the total cross-section $\gamma^* - dipole (\xb_{01})$ and for
 the
 related structure function (we assume $R = F_L/F_T$ to be small in the
 relation between the total cross-section and the structure function)
\beq
\label{gonium2}
\frac{Q^2} {4 \pi^2 \alpha_{e.m}} \sigma_{\gamma^*}^{d} (x_{bj},Q^2;x_{01}^2)
 = 
 F_{\gamma}^{d}(x_{bj},Q^2;x_{01}^2)
= \frac{2 \alpha_s N_c}{\pi} \int^{ }_{ } \frac{d \gamma}{2 i \pi}
 (Q^2 x_{01}^2)^{\gamma} 
 h(\gamma) \frac{v(\gamma)}{\gamma}  
e^{\frac{\alpha_s N_c}{\pi} \chi(\gamma) \ln\frac{1}{x_{bj}}}.
\eq
The dipole $(\xb_{01})$ being part of a bound state, for example extracted
from an onium, 
one has now to average with respect to the coresponding wave function,
 $\Phi^{(0)}(\underline{x}_{01},z_1).$ The initial dipole state is supposed to
 be well localized in transverse space. Namelly,
 its transverse size, given by the scale $M^2$ which is defined via
\beq
\label{defM}
(M^2)^{-\gamma} = \int d^2\underline{x}_{01} (x_{01}^2)^{\gamma} \, dz_1
 \Phi^{(0)}(\underline{x}_{01},z_1),
\eq
is assumed to be {\it perturbative.}
One then obtains for the onium structure function
\beqa
\label{crosshadronnonpert}
F^{{\cal O}nium}(x_{bj},Q^2;M^2) = 2 \frac{\alpha_s N_c}{\pi} \int^{ }_{ }
 \frac{d \gamma}{2 i \pi}  
 h(\gamma) \frac{v(\gamma)}{\gamma}  \left(\frac{Q^2}{M^2}\right)^{\gamma} 
\exp\left(\frac{2\alpha_s N_c}{\pi} \chi(\gamma) \ln\frac{1}{x_{bj}}\right).
\eqa
It is then possible to apply formula (\ref{crosshadronnonpert}) to peculiar 
structure
 functions, namely
\beq
\label{predgen}
\left(\begin{array}{c}
F_T \\ F_L \\ F_G \end{array} \right) 
= \frac{2\alpha_s N_c}{\pi} \int^{ }_{ }  \frac{d \gamma}{2 i \pi}  
  \left(\frac{Q^2}{M^2}\right) ^{\gamma} 
\exp\left(\displaystyle \frac{2\alpha_s N_c}{\pi} \chi(\gamma) \ln\frac{1}
{x_{bj}}
\right) \left(\begin{array}{c}
h_T \\ h_L \\ 1 \end{array} \right)
\frac{v(\gamma)}{\gamma} 
\eq
where $F_{T(L)}$ is the structure function corresponding to transverse 
(longitudinal)
 photons and $F_G$ the gluon structure function.
The coefficient functions 
\beq
\label{defh}
\left(\begin{array}{c}
h_T \\ h_L \end{array} \right) = \frac{\alpha_s}{ 3 \pi \gamma} 
\frac{(\Gamma(1 - \gamma) \Gamma(1 + \gamma))^3}{\Gamma(2 - 2\gamma) \Gamma(2 +
 2\gamma)} \frac{1}{1 - \frac{2}{3} \gamma} \left( \begin{array}{c} (1 +
 \gamma)
(1 - \frac{\gamma}{2}) \\ \gamma(1 - \gamma) \end{array} \right)
\eq
were computed in ref \cite{catani}. 
\noindent
When $x_{bj}$ is small, the $\gamma-$integration can be performed by the
 steepest-descent method. 
The corresponding
asymptotic saddle point is located at
 $\gamma = \de$, which defines the BFKL anomalous dimension.
Expanding the $\chi$ function
 around $\de,$
we obtain a saddle point at  
\beqa
\label{saddle}
\gamma_s &=& \frac{1}{2} \left( 1 - a \ln \frac{Q}{Q_0}\right), \quad
\mbox {where} \quad  a  = \left(\frac{\bar{\alpha} N_c}{\pi} 7 \zeta(3)
 \ln\frac{1}{x_{bj}}\right) ^{-1}.
\eqa
The approximation of expanding $\chi(\gamma)$ around $\de$ is valid when 
\beq
\label{applicabilite}
a \ln\left(\frac{Q}{M}\right) \simeq \ln \frac{Q}{M}/\ln\frac{1}{x_{bj}}
 \ll 1,
\eq
that is the small $x_{bj}$, moderate $Q/M$ kinematical domain.
This yields
\beq
\label{predF2}
F_2 \equiv F_T + F_L = C a^{1/2} \frac{Q}{M} \exp \left( (\alpha_{P} -1)
 \ln\frac{1}{x_{bj}} - \frac{a}{2} \ln^2 \frac{Q}{Q_0} \right),
\eq
where $\alpha_{P}$ is defined by Eq. (\ref{alphap}).
Thus, $F_2$ depends only on 3 parameters, $C, \ M \  {\rm and} \  
\alpha_{P}.$
Suppose we can fit $F_2$ with this form. Then,  get a prediction for $F_G$ and 
$R = F_L / F_T$ without any free parameter.
Namely,
\beq
\label{fgf2}
\frac{F_G}{F_2} = \left. \frac1{h_T + h_L}\right|_{\gamma = \gamma_s}\equiv
 \frac{3 \pi \gamma_s}{\alpha_s} 
\frac{1 - \frac{2}{3}\gamma_s}{1 + \frac{3 }{2}\gamma_s - \frac{3}{2}
\gamma_s^2}
 \frac{\Gamma(2 - 2 \gamma_s) \Gamma(2 + 2 \gamma_s)}{(\Gamma(1 - \gamma_s)
\Gamma(1 + \gamma_s))^3}  
\eq
and
\beq
\label{R}
R = \frac{h_L}{h_T}(\gamma_s) = \frac{\gamma_s (1 - \gamma_s)}{(1 + \gamma_s)
(1 -
 \frac{\gamma_s}{2})},
\eq
where  $\gamma_s$ is given  by the expression (\ref{saddle}).
Note that the overall non-perturbative normalization $C$ does not enter $R$ 
and $F_G/F_2$.
It is possible to apply this analysis to the proton \cite{nprw}. This requires
 some additional asumptions when considering the coupling of the dipole
cascade to the proton.
 It leads to a successful description of the HERA data \cite{h1}. It also
 provides a
 prediction for the gluon density
based on the BFKL dynamics, and a prediction for the ratio $R,$ using formulas
(\ref{fgf2}) and (\ref{R}). All the previous annalysis was done by performing
an expansion around the small $x_{bj}$ behaviour of the BFKL Pomeron, that is
by considering $\ln 1/x_{bj}$ as a big parameter, which leads to a saddle
point close to $\gamma = \de$ (see Eq. (\ref{saddle})). A different
 analysis of the modification, due to the BFKL dynamics, 
of the anomalous dimension of the double 
logarithmic approximation common to DGLAP and BFKL
 can be done by considering now $\ln Q^2/M^2$
 as a big parameter. The corresponding saddle point of Eq. (\ref{predgen})
is now around $\gamma=0.$
In Mellin space, 
the double logarithmic expression of the anomalous 
dimension is given by $\gamma_{\omega}(\alpha_s) = 3\alpha_s/(\pi \omega),$
and expansion of the $\chi$ function (\ref{chi}) around $\gamma =0$
leads to corrections given by powers of 
$3\alpha_s/(\pi \omega).$
This method provides an extension of the
 domain
 of applicability (\ref{applicabilite})
\cite{nprw2}.

 As it has been seen in this section, the dipole model
can be safely applied when the two scales of the process are both perturbative,
as it is the case for $e^{\pm}-onium$ scattering. The application to
$e^{\pm}-p$ scattering requires some assumptions for the coupling to the 
proton. Because of the well-known diffusion in transverse momentum space,
such an application of the dipole model, although successful
 \cite{npr,nprw,wal7}, 
cannot  be considered as a clean test of high-energy perturbative regge 
dynamics. A possible test of such dynamics could be based on single jets 
events
in DIS \cite{jet1} or double jets events in hadron-hadron collision 
\cite{mulnav}.
Another interesting test of BFKL dynamics would be the $\gamma^*-\gamma^*$
events in $e^+-e^-$ colliders at high energy in the center of mass of the
virtual photon pair and with high (perturbative) photons virtualities.
This has been already proposed in the framework of the original BFKL
equation \cite{gammagamma}.  
Such a process can equivalently be described 
in the dipole picture of BFKL dynamics. This will be developped elsewhere 
\cite{rrw}.

\section{Conclusion}
In this article we have shown the exact equivalence between BFKL and dipole
approaches for the onium-onium cross-section at fixed impact parameter. This
 proof
relies on conformal properties of the dipole cascade and of the elementary 
dipole-dipole cross section. We have also obtained asymptotic expressions
 for
 the
 dipole distribution inside an onium and for the  onium-onium cross-section at
 fixed impact parameter. These results agrees with previous 
numerical
simulations.  We also apply the dipole model to onium-$e^{\pm}$ deep inelastic
scattering, using the $k_T$-factorization, and obtain 
 predictions for various structure functions in the BFKL dynamics.
The different elementary 
cross-sections
used in this paper are computed using eikonal techniques. 

Relying on the same
 conformal
properties, it should be possible to get analytical
expressions for the multipomeronic contributions to the onium-onium 
cross-section,
which are expected to be important for large rapidities.

From a phenomenological point of vue, the dipole framework could be applied
to other inclusive processes. The application of this technique for 
exclusive quantities remains however an open question, due to the use
of light-cone quantization, in which the intermediate states are unphysical.
\\

\noindent
{\bf Acknowledgements:}

We thank Robi Peschanski for discussions, and Andr\'e Morel for a careful
reading of the manuscript and many fruitful comments.
 S.W wishes to thank Hans Lotter, 
Jochen Bartels and Gregory Korchemsky for comments. He warmly thanks
Al Mueller for many illuminating discussions in Columbia and in Orsay.
He thanks the Alexander von Humboldt Foundation and 
the II. Instit\"ut f\"ur Theoretische Physik at DESY for support.

\eject
\begin{appendix}
\section{Appendices}
\subsection{Calculation of $\sigma_{DD}$ using eikonal methods}
\label{calculsigmadd}
\setcounter{equation}{0}

In this appendix we compute the dipole-dipole cross-section using eikonal
 techniques. 
In QCD, the eikonal current due to a fast quark of momentum $p$, responsible 
for
 the emission of a soft gluon of momentum $k$ ($k \ll p$) and color $a$ reads
\beq
\label{courant1}
j^{\nu}(k)= -i g T^a \frac{p^{\nu}}{p.k+ i \epsilon}.
\eq
Let us consider the scattering of two dipoles.
 Let $\underline{x}_0$ ($\underline{x}_1$) be the transverse coordinate of
 the free quark (antiquark) making the right moving dipole and
 $\underline{x}_0'$ ($\underline{x}_1'$) the coordinates of the corresponding
quark (antiquark) making the left moving dipole.
In this mixed representation where the fast radiating particule is represented
 in impact parameter space, and the radiated gluon is represented in momentum
 space, the eikonal current corresponding to a quark of transverse coordinate
 $\underline{x}$ then reads
\beq
\label{courant2}
j^{a\nu}(k)= -i g T^a \frac{\bar{\eta}^{\nu}}{\bar{\eta}.k+ i \epsilon} 
e^{-\displaystyle i\underline{x}.\underline{k}},
\eq
where $\bar{\eta} = (\stackrel+1,\stackrel-0,\stackrel\bot0).$
This current is responsible for a term $-i j^{a\nu}(k) \epsilon_{\nu}$ when
 computing the amplitude of emission of a gluon by the quark. This terms arises
when evaluating the evolution operator $T\,\exp{-i \int d^4 x j^{a\nu}(x)
 A_{a\nu}(x)}$ due to the usual hamiltonian $\int d^3x \,j^{a\nu}(x) A_{a\nu}
(x).$ 
Note that  this current is completely described by the transverse size of the 
dipole and does not require any additional information
 about its internal structure.
 
\noindent
The eikonal current corresponding to the right moving dipole then reads
\beq
\label{courant3}
j^{a\nu}(k)= -i g T^a \frac{\bar \eta^{\nu}}{\bar \eta.k+ i \epsilon}( 
e^{-\displaystyle i\underline{x}_0.\underline{k}}- e^{-\displaystyle i
\underline{x}_1.\underline{k}}),
\eq
and a corresponding formula for the left moving dipole, replacing $\bar \eta$
 by $\eta=(\stackrel+0,\stackrel-1,\stackrel\bot0).$

\noindent
Let us now compute the graph $A_1$ represented in figure \ref{sigmadd}.
\begin{figure}[htb]
\centering
\epsfysize=5.0cm{\centerline{\epsfbox{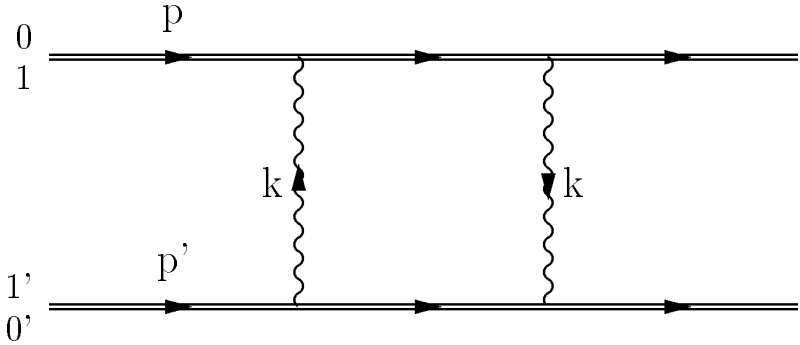}}}
\caption{Contribution to the dipole-dipole scattering.}
\label{sigmadd}
\end{figure}
It reads, in Feynman covariant gauge,
\beqa
\label{A1}
&&\hspace{-1cm}A_1 = \frac{1}{N_c^2}\sum_{ab} \frac{1}{2 p^+} 
 \frac{1}{2 p^{'-}}\int \frac{d^4k}{(2 \pi)^4}  \left\{\frac{2(-p.k)}{i}(-i
 j^{b+}
(-k))\right\} \left\{ \frac{2(p'.k)}{i}  (-i j^{b-}(k))\right\}\nonumber\\
&&\hspace{-1cm}\times \, (-i j^{b+}(k))(-i j^{b-}(-k))\frac{(-i)^2}{(k^2)^2}.
\eqa
The terms $\frac{1}{2 p^+}  \frac{1}{2 p^-}$ are related to the normalisation
 of
the initial dipole states and $\frac{2(-p.k)}{i}$ and $\frac{2(p'.k)}{i}$ are
 due
 to the fact that the considered amplitude is computed for amputed propagators.
The color factor reads
\beq
\label{couleur}
\frac{1}{N_c^2}\sum_{ab}Tr\, T^aT^b Tr \,T^aT^b=\frac{N_c^2-1}{4 N_c^2},
\eq
which equals $\frac{1}{4}$ in the large $N_c$ limit.

\noindent
Using the expression (\ref{courant3}), this yields
\beqa
\label{A2}
&&\hspace{-1cm} A_1 = \frac{g^4}{4} \int \frac{d^4l}{(2 \pi)^4}\frac{1}
{(k^2)^2}\,
 \frac{1}{k^- +i \epsilon} \, \frac{1}{k^+ -i \epsilon}\nonumber\\
&&\hspace{-1cm}\times \, ( e^{\displaystyle i\underline{x}_0 . \underline{k}}-
 e^{\displaystyle i\underline{x}_1 .\underline{k}}) \, ( e^{-\displaystyle i
\underline{x}_0 . \underline{k}}- e^{-\displaystyle i\underline{x}_1 . 
\underline{k}})\, 
( e^{-\displaystyle i\underline{x}_0'.\underline{k}}- e^{-\displaystyle i
\underline{x}_1'.\underline{k}})\, ( e^{\displaystyle i\underline{x}_0'.
\underline{k}}- e^{\displaystyle i\underline{x}_1'.\underline{k}}).
\eqa
One has also to consider the crossed diagram $A_2$. In order to integrate with 
respect to $k^-$ and $k^+$ we consider two equivalent representations of $A_1$
 and $A_2$, obtained by changing the sign of $k$ (see figure \ref{tousgraphes}.
\begin{figure}[htb]
\begin{picture}(500,220)(0,0)
\put(-130,130){\epsfysize=3.8cm{\centerline{\epsfbox{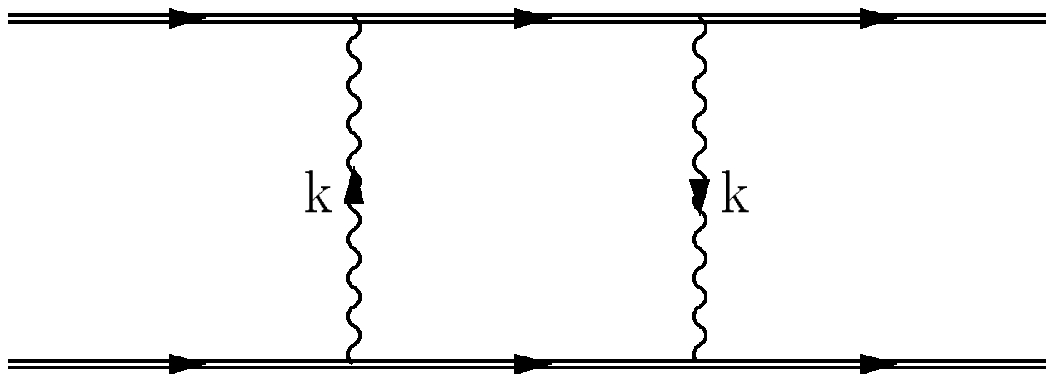}}}}
\put(105,120){$A_1$}
\put(130,130){\epsfysize=3.8cm{\centerline{\epsfbox{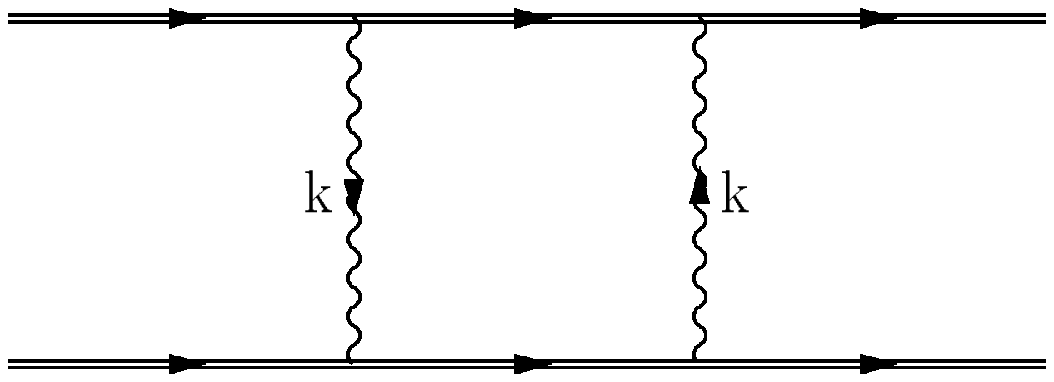}}}}
\put(370,120){$A_1$}
\put(-130,10){\epsfysize=3.8cm{\centerline{\epsfbox{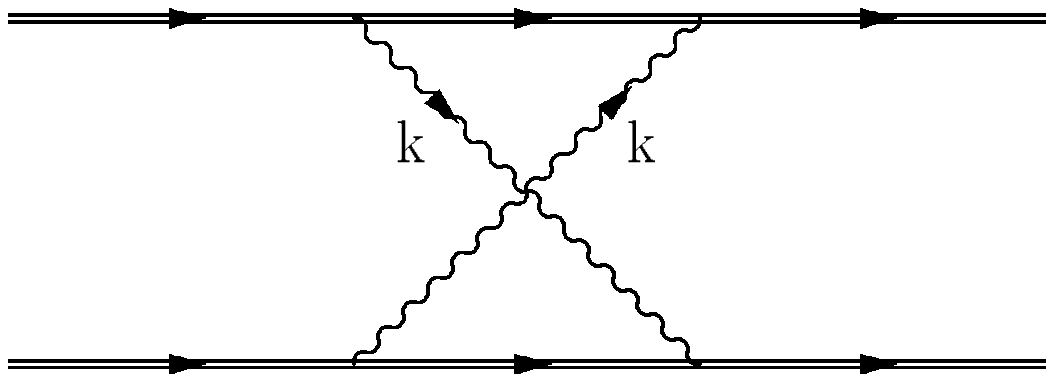}}}}
\put(105,0){$A_2$}
\put(130,10){\epsfysize=3.8cm{\centerline{\epsfbox{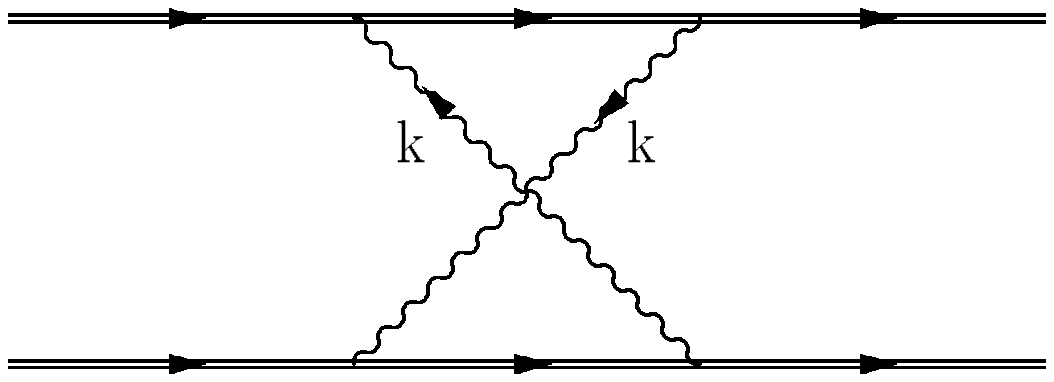}}}}
\put(370,0){$A_2$}
\end{picture}
\caption{Contributions to the elementary dipole-dipole scattering.}
\label{tousgraphes}
\end{figure}
Forgetting the phase factor for a moment, the forward amplitude $A$
then reads
\beq
\label{Atotal1}
A = A_1 + A_2 = \frac{g^4}{4} \int \frac{d^2 \underline{k}}{(2 \pi)^2}
\frac{dk^-}{2\pi}\frac{dk^+}{2\pi}\frac{1}{(k^2)^2} \frac{1}{2}\left(
\frac{1}{k^-+i \epsilon}+\frac{1}{-k^-+i \epsilon}\right)\, \left(\frac{1}
{k^++i
 \epsilon}+\frac{1}{-k^++i \epsilon}\right).
\eq
The two last terms reads $(-2\pi i\delta(k^-))(-2\pi i\delta(k^+)),$
which finally yield, performing the integration with respect to $k^-$ and $k^+$
\beq
\label{Atotal2}
A = -\frac{\alpha_s^2}{2}\int \frac{d^2 \underline{k}}{(\underline{k}^2)^2}
\left(2-e^{\displaystyle i\underline{k}.\underline{x}_{01}}-e^{-\displaystyle i
\underline{k}.\underline{x}_{01}}\right)\left(2-e^{\displaystyle i
\underline{k}.
\underline{x}'_{01}}-e^{-\displaystyle i\underline{k}.\underline{x}'_{01}}
\right),
\eq
which was obtained in Ref. \cite{muellerpatel} by computing elementary Feynman
 diagrams.
If we neglect the dependence of  $\Phi(x_{01},z)$ and $n(x_{01},x,Y)$ with
respect to the dipoles orientation (see appendix \ref{F1approche}), 
 it is possible to average with respect to the angle of these
 dipoles.
$A$ then reads
\beq
\label{Atotal3}
A = -4 \pi \alpha_s^2\int^{\infty}_0 \frac{dk}{k^3}[1-J_0(kx_{01})]\,
[1-J_0(kx'_{01})]
\eq
Using the identity
\beq
\label{JJ}
\int^{\infty}_0 \frac{dk}{k^3}[1-J_0(kx_{01})]\,[1-J_0(kx'_{01})] =
 \frac{1}{4} x_<^2\left[1+ \ln \frac{x_>}{x_<}\right],
\eq
where $x_< = Min(x_{01},x'_{01})$ and $x_> = Max(x_{01},x'_{01}),$
this finally yields for the corresponding cross-section
\beq
\label{Atotal4}
\sigma_{DD}(x_{01},x'_{01})= -2 A = 2 \pi \alpha_s^2 x_<^2\left[1+ \ln
 \frac{x_>}{x_<}\right].
\eq
This average result is however sufficient when computing the {\it
 total} cross-section (see appendix \ref{F1approche}).

Another representation of this elementary cross-section is very useful.
Consider 
\beq
\label{repsigmadd1}
I = \frac{\alpha_s^2}{2}\int^{+\infty}_{-\infty}\frac{d \nu}{(\nu^2 +
 \frac{1}{4})^2} x_1^{1+2 i \nu} x_2^{1-2 i \nu}.
\eq
When $x_1 > x_2$ $(x_1 < x_2)$ one can close the integration contour around
$+i \infty$ $(-i \infty)$ , so that one pick up the (simple) pole at $\nu =
 i/2$ $(\nu = -i/2)$ .
Thus,
\beq
\label{repsigmadd2}
I = \frac{\alpha_s^2}{2} 2 i \pi \frac{d}{d\nu} \left.\frac{e^{2 i \nu \ln
 \frac{x_>}{x_<}}}{\left(\nu \pm \frac{i}{2}\right)^2}\right|_{\nu=\pm 
\frac{i}{2}} x_> x_<,
\eq
where $x_<\,=Min(x_1,x_2)$ et $x_> \, = Max(x_1,x_2)$.
This finally yields the expected result:
\beq    
\label{repsigmadd}
\sigma_{DD}(x'_1,x'_2)= \frac{\alpha_s^2}{2}\int^{+\infty}_{-\infty}
\frac{d \nu}{(\nu^2 + \frac{1}{4})^2} (x'_1)^{1+2 i \nu} (x'_2)^{1-2 i \nu}.
\eq
Note that this can be written equivalently as (see the following section)
\beq
\label{repsigmaddEq}
\sigma_{DD}(x'_1,x'_2)= \frac{\alpha_s^2 \, x'_1 \, x'_2}{4} 
\int^{+\infty}_{-\infty}\frac{d \nu}{(\nu^2 + \frac{1}{4})^2}
 \lim_{q \to 0} \left[E^{0\nu*}_{q}(x'_1)\,E^{0\nu}_{q}(x'_2)\right].
\eq
\subsection{Calculation of $\sigma_{DD}(\underline{x},
\underline{x}',\underline{b}-\underline{b}')$}
\label{calculsigmaddq}
In this appendix we compute the elementary dipole-dipole cross-section at
 fixed impact
parameter. As in the appendix \ref{calculsigmadd}, we consider two dipoles
 of transverse
sizes $\underline{x} =  \underline{x}_0 - \underline{x}_1$ and 
 $\underline{x}' =  \underline{x}'_1 - \underline{x}'_0,$ whose centers are
 situated at $\underline{b} = \frac{\underline{x}_0 + \underline{x}_1}{2}$
 and $\underline{b}' = \frac{\underline{x}'_0 + \underline{x}'_1}{2}.$

Let us compute the non-forward scattering amplitude of these two dipoles. In
the high-energy limit $-t<< s=2p.p',$ the exchanged momentum is
 quasi-transverse
\beq
\label{qtransverse}
q \simeq (\stackrel+0,\stackrel-0,\stackrel{\bot}{\underline{q}}).
\eq
The graphs to be computed are displayed in  figure \ref{tousgraphesq}.
\begin{figure}[htb]
\begin{picture}(500,220)(0,0)
\put(-130,130){\epsfysize=3.8cm{\centerline{\epsfbox{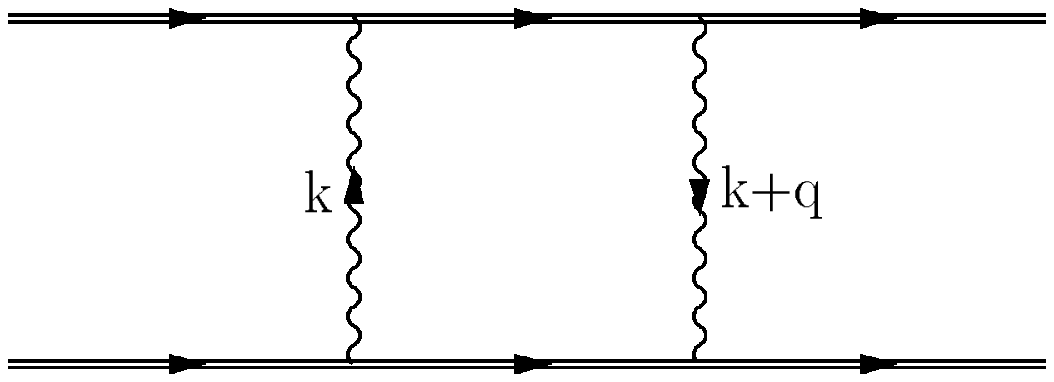}}}}
\put(105,120){$A_1$}
\put(130,130){\epsfysize=3.8cm{\centerline{\epsfbox{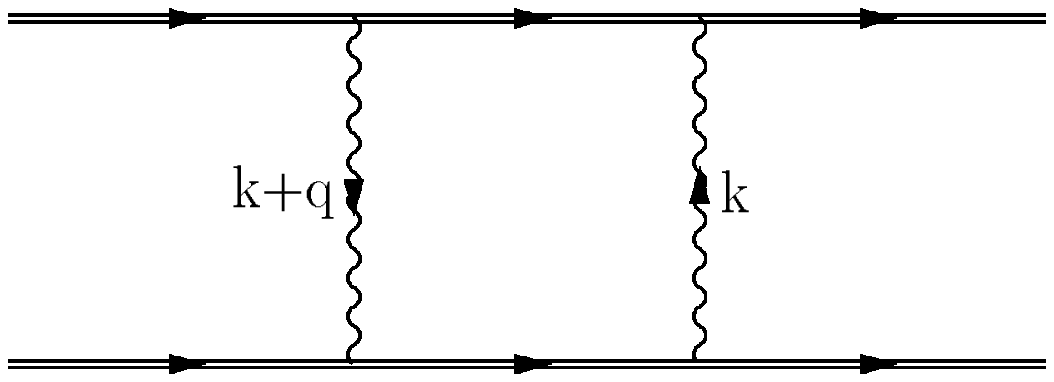}}}}
\put(370,120){$A_1$}
\put(-130,10){\epsfysize=3.8cm{\centerline{\epsfbox{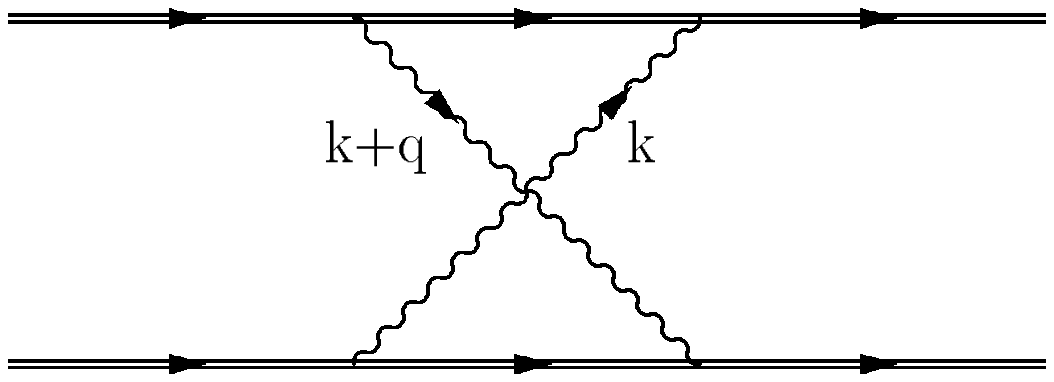}}}}
\put(105,0){$A_2$}
\put(130,10){\epsfysize=3.8cm{\centerline{\epsfbox{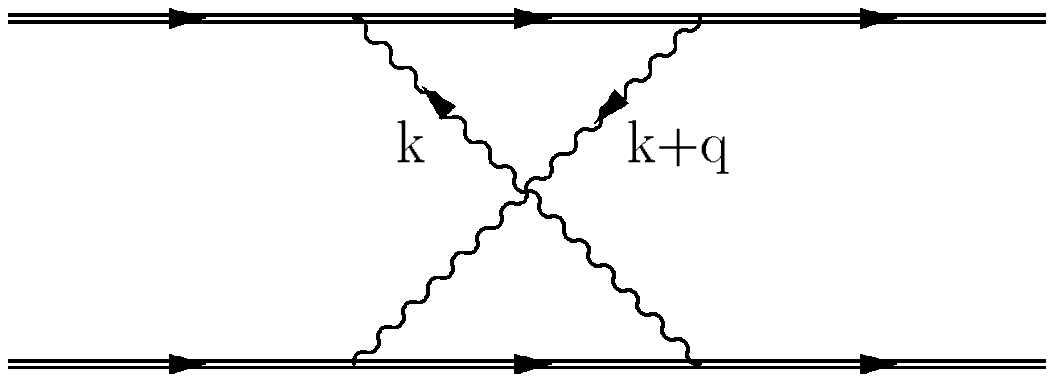}}}}
\put(370,0){$A_2$}
\end{picture}
\caption{Contributions to the elementary dipole-dipole scattering with non
 zero exchanged momentum.}
\label{tousgraphesq}
\end{figure}
These graphs are computed using the rules of appendix \ref{calculsigmadd}.
 Since $q^- = q^+ =0,$ the expression for the corresponding amplitude is
 similar to the one obtained in equation (\ref{Atotal2}), except for phase
 factors. It reads
\beqa
\label{Atotalq}
&&A = -\frac{\alpha_s^2}{2} \int \frac{d^2 \underline{q}}{(2 \pi)^2} \int
 \frac{d^2 \underline{k}}{\underline{k}^2 (\underline{k} +\underline{q})^2}
\left(e^{\displaystyle i \underline{x}_0.\underline{k}} - e^{\displaystyle i
 \underline{x}_1.\underline{k}}\right)\left(e^{-\displaystyle i
 \underline{x}_0.
(\underline{k}+\underline{q})} - e^{-\displaystyle i \underline{x}_1.
(\underline{k}+\underline{q})}\right) \nonumber \\
&& \hspace{2 cm}\times \, 
\left(e^{-\displaystyle i \underline{x}'_0.\underline{k}} -
 e^{-\displaystyle i \underline{x}'_1.\underline{k}}\right)
\left(e^{\displaystyle i \underline{x}'_0.(\underline{k}+\underline{q})}
 - e^{\displaystyle i \underline{x}'_1.
(\underline{k}+\underline{q})}\right).
\eqa
Note that this expression is overall translationally invariant. However, 
it will 
depend on the relative distance between the two dipoles and on their angles.
Expanding the previous expression and integrating with respect to the angles
 of $\underline{k}$ and  $\underline{k}+\underline{q},$ one obtains several
 Bessel functions
\beq
\label{Aq1}
A =  -\frac{\alpha_s^2}{2} \left \{ I(|\underline{x}_0 - \underline{x}_{0'}|)
-I(|
\underline{x}_0 - \underline{x}_{1'}|)-I(|\underline{x}_1 - 
\underline{x}_{0'}|)+
I(|\underline{x}_1 - \underline{x}_{1'}|\right \}^2,
\eq
where
\beq
\label{defI}
I(\underline{x}) =  \int^{+\infty}_{\rho} \frac{d k}{k} \, J_0(k x).
\eq
$\rho$ is an infrared cut-off which regularize the divergency at $k =0.$
In the limit $\rho \to 0,$ $I$ can be computed (see Ref. \cite{mueller94})
\beq
\label{calculI}
I =  \lim_{\lambda \to 0} \left[ \int^{+\infty}_0 \frac{d k}{k^{1 - \lambda}} 
J_0(kx) - \int^{\rho}_0 \frac{d k}{k^{1-\lambda}} J_0(kx) \right]
= \psi(1) + \ln 2 - \ln x - \ln \rho.
\eq
When evaluating $A,$ the constant $\psi(1) + \ln 2$ and the infrared divergent
 term $\ln \rho$ cancels. $A$ then reads
\beq
\label{calculA}
A =  -\frac{\alpha_s^2}{2}  \left\{ \ln \frac{|\underline{x}_0 - 
\underline{x}_{0'}| |\underline{x}_1 - \underline{x}_{1'}|} {|\underline{x}_0 -
 \underline{x}_{1'}| |\underline{x}_1 - \underline{x}_{0'}|} \right\}^2
\eq
Defining 
\beq
\label{defbb}
\underline{b} = \frac{\underline{x}_0 + \underline{x}_1}{2} \quad \mbox{ and
} \quad \underline{b}' = \frac{\underline{x}_{0'} + \underline{x}_{1'}}{2},
\eq
the non-forward cross-section finally reads
\beq
\label{sigmaddb}
\sigma_{DD}(\underline{x},
\underline{x}',\underline{b}-\underline{b}') = - 2 A
 =
\alpha_s^2 \,  \left\{ \ln \frac{|\underline{b}' - \underline{b} + \frac{
\underline{x} + \underline{x}'}{2}||\underline{b} - \underline{b}' + \frac{
\underline{x} + \underline{x}'}{2}| }  {|\underline{b}' - \underline{b} + 
\frac{
\underline{x} - \underline{x}'}{2}||\underline{b} - \underline{b}' + \frac{
\underline{x} - \underline{x}'}{2}| }\right\}^2,
\eq
as quoted in Ref. \cite{salam}. Note that it only depends on the relative
 distance $\underline{b} - \underline{b}'$ because of the overall
 translational 
invariance. This expression can also be obtained
by evaluating this scattering amplitude in the laboratory frame of one of
the two onia. In this frame, this amplitude is expressed as an 
 eikonal phase (computed in terms of a  Wilson-loop), due to the change of
 the wave
function of the slow moving onium in the color field of the fast moving onium
 \cite{kmw}.
Let us now show that this cross-section can be expanded on the basis of the 
functions
$E^{n,\nu}.$
In the standard Regge calculation, instead of considering the scattering of two
 dipoles,
one considers the scattering of two gluons of momenta $-\underline{k}$ and
 $\underline{k} + \underline{q}$ (see figure \ref{gluongluon}).
\begin{figure}[htb]
\begin{picture}(500,220)(0,0)
\put(-130,130){\epsfysize=3.8cm{\centerline{\epsfbox{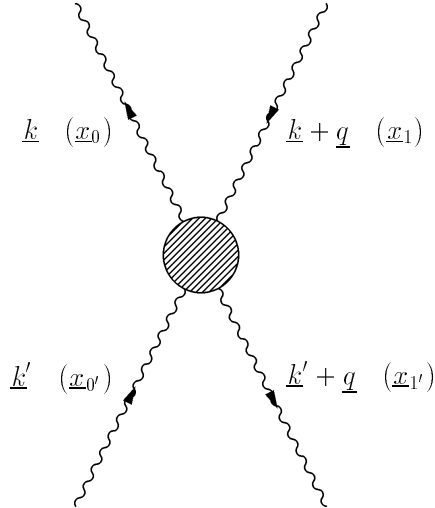}}}}
\end{picture}
\caption{Gluon-Gluon diffusion in the Regge framework.}
\label{lipatov0}
\end{figure}
In the lowest order approximation, the contribution of this graph is
\beq
\label{gluongluon}
A_{gg} = \frac{\delta^2(\underline{k} - \underline{k}')}{\underline{k}^2
 (\underline{k}+\underline{q})^2},
\eq
which gives, after Fourier transform in impact parameter space \cite{liprev}
\beq
\label{defa}
a(\underline{x}_0,\underline{x}_1,\underline{x}_{0'},\underline{x}_{1'}) = 
(2 \pi)^2 \ln (|\underline{x}_{00'}|\lambda) \ln (|\underline{x}_{11'}|
\lambda),
\eq
where $\lambda$ is the mass of the gluon introduced in order to remove the
 infrared 
divergency.
Since one is interested in the coupling with color neutral states, this 
expression can equivalently be replaced by
\beq
\label{a1}
a(\underline{x}_0,\underline{x}_1,\underline{x}_{0'},\underline{x}_{1'}) = 2
 \pi^2
 \ln \left|\frac{\underline{x}_{00'} \underline{x}_{11'}}{\underline{x}_{01'}
\underline{x}_{0'1}}\right| \, \ln \left|\frac{\underline{x}_{00'}
 \underline{x}_{11'}}{\underline{x}_{01}\underline{x}_{0'1'}}\right|
\eq
because of the conservation of color current \cite{liprev}.
In the case of the dipole model, instead of labeling the gluons, one labels the
 dipoles.
Thus, the contributions of the figure \ref{tousgraphesq} should be equal to the
 contribution of figure \ref{lipatov0} plus permutations $(\underline{x}_0 
\leftrightarrow \underline{x}_1)$ and $(\underline{x}_{0'}\leftrightarrow
 \underline{x}_{1'})$ (except for 
normalisation factors).
Indeed, one can check that
\beqa
\label{equivalence}
&&\sigma_{DD}(\underline{x},
\underline{x}',\underline{b}-\underline{b}')=
 \frac{\alpha_s}{(2 \pi)^2} [a(\underline{x}_0,\underline{x}_1,
\underline{x}_{0'},
\underline{x}_{1'}) + (\underline{x}_0 \leftrightarrow \underline{x}_1) +
 (\underline{x}_{0'}\leftrightarrow \underline{x}_{1'}) + (\underline{x}_0 
\leftrightarrow \underline{x}_1 , \underline{x}_{0'}\leftrightarrow 
\underline{x}_{1'})] \nonumber \\
&&= 2 \alpha_s [a(\underline{x}_0,\underline{x}_1,\underline{x}_{0'},
\underline{x}_{1'}) + (\underline{x}_0 \leftrightarrow \underline{x}_1)].
\eqa
Using the expression (26) of Ref. \cite{lipatov86} in the Born approximation,
that is making  $g=0$, one finally obtains
\beqa
\label{devsigma}
&&\sigma_{DD}(\underline{x},
\underline{x}',\underline{b}-\underline{b}')= 
\frac{2 \alpha_s^2}{(2 \pi)^2} \sum_{n=-\infty}^{+\infty} \int_{-\infty}^{+
\infty}
 d \nu  \int d^2 \underline{w} \left(\nu^2 + \frac{n^2}{4}\right) \frac{1 +
 (-1)^n}{\left( \nu^2 + \left(\frac{n-1}{2} \right)^2 \right) \left( \nu^2 + 
\left ( \frac{n+1}{2} \right)^2 \right)} \nonumber \\
&&\times \, E^{n,\nu*}\left(\underline{b} + \frac{\underline{x}}{2} - 
\underline{w},\underline{b} - \frac{\underline{x}}{2} - \underline{w}\right) \,
 E^{n,\nu}\left(\underline{b}' + \frac{\underline{x}'}{2} - \underline{w},
\underline{b}' - \frac{\underline{x}'}{2} - \underline{w}\right),
\eqa
since 
\beq
\label{opposeE}
E^{n,\nu}(x_{b0},x_{a0}) = (-1)^n E^{n,\nu}(x_{a0},x_{b0}).
\eq

One can check on this expression that when integrating over $\underline{b}' -
 \underline{b} $ and averaging over angles, one recovers the total
cross-section
(\ref{repsigmadd}).
Indeed, using the mixed representation (\ref{Eq1}), one obtains
\beqa
\label{devsigmaq}
&&\sigma_{DD}(\underline{x},
\underline{x}',\underline{b}-\underline{b}')= 2 
\alpha_s^2 \int \frac{d^2 \underline{q}}{(2 \pi)^2}\,  e^{\displaystyle - i 
\underline{q}.(\underline{b}' - \underline{b})} \, \frac{x \, x'}{16} 
\sum_{n=-\infty}^{+\infty} \int_{-\infty}^{+\infty} d \nu \,  \frac{1 + (-1)^n}
{\left( \nu^2 + \left(\frac{n-1}{2} \right)^2 \right) \left( \nu^2 + \left ( 
\frac{n+1}{2} \right)^2 \right)} \nonumber \\
&&\hspace{3cm} \times \, E^{n,\nu*}_q(x') E^{n,\nu}_q(x).
\eqa
Thus,
\beqa
\label{sigmabsigma}
&&\hspace{-1.5cm} \sigma_{DD}(x,x') = \int d^2 (\underline{b}' - \underline{b})
 \,  \sigma_{DD} (\underline{x},
\underline{x}',\underline{b}-\underline{b}')
\nonumber \\
&&\hspace{-.5cm}= 2 \alpha_s^2 \,  \frac{x x'}{16} \sum_{n=-\infty}^{+\infty} 
\int_{-\infty}^{+\infty} d \nu \,  \frac{1 + (-1)^n}{\left( \nu^2 + \left(
\frac{n-1}{2} \right)^2 \right) \left( \nu^2 + \left ( \frac{n-1}{2} \right)^2
 \right)} \lim_{q \to 0} \left[E^{n,\nu*}_q(x') E^{n,\nu}_q(x)\right].
\eqa
Using the expansion (A.1) of Ref. \cite{lipatov86}, one obtains
\beqa
\label{devEq}
&&\hspace{-.7cm}\left. E^{n,\nu*}_q(x') E^{n,\nu}_q(x)\right|_{q \ll 1/x,1/
{x'}}
 \nonumber \\
&&\hspace{-.6cm}= \left(\frac{x x^{'*}}{x^* x'}\right)^{n/2} \left|\frac{x}{x'}
\right|^{-2 i \nu} \left[1 + \left(\frac{q x^*}{q^* x}\right)^n |q x|^{4 i \nu}
 e ^{i \delta(n,\nu)} \right]
\left[1 + \left(\frac{q^* x'}{q x^{'*}}\right)^n |q x'|^{-4 i \nu} e ^{-i 
\delta(n,\nu)} \right] \nonumber \\
&&\hspace{-.6cm} = \left(\frac{x x^{'*}}{x^* x'}\right)^{n/2} \left|\frac{x}
{x'}
\right|^{-2 i \nu} \left[1 + \left(\frac{x' x^*}{x x^{'*}}\right)^n \left|
\frac{x}
{x'}\right|^{4 i \nu}
+ \left(\frac{q x^*}{q^* x} \right)^n |q x|^{4 i \nu} e ^{i \delta(n,\nu)} + 
\left(\frac{q^* x'}{q x^{'*}} \right)^n |q x'|^{-4 i \nu} e ^{-i \delta(n,\nu)}
\right] \nonumber \\
\eqa
where $e^{i \delta(n,\nu)}$ is a phase given by Eq. (A.2) of Ref.
 \cite{lipatov86}.
When $q \to 0,$ only the two first terms remain. Due to the symetry of 
$\frac{1 +
 (-1)^n}{\left( \nu^2 + \left(\frac{n-1}{2} \right)^2 \right) \left( \nu^2 +
 \left
 ( \frac{n-1}{2} \right)^2 \right)}$ when $\nu \to -\nu$ or  $n \to -n,$
the contribution of these two terms is the same when summing over $n$ and 
integrating over $\nu.$ 
Thus,  
\beq
\label{sigmaangle}
\sigma_{DD}(x,x')=  4 \alpha_s^2 \frac{|x| |x'|}{16} \sum_{n=-\infty}^{+\infty}
 \int_{-\infty}^{+\infty} d \nu \,  \frac{1 + (-1)^n}{\left( \nu^2 + \left(
\frac{n-1}{2} \right)^2 \right) \left( \nu^2 + \left ( \frac{n+1}{2} \right)^2
 \right)}
\left(\frac{x x^{'*}}{x^* x^{'}}\right)^{n/2} \left|\frac{x}{x'}\right|^{-2 i
 \nu}.
\eq
Averaging over the angles, the only remaining term is $n=0.$
Thus,
\beq
\label{resultatequiv}
\sigma_{DD}(x,x') = \frac{\alpha_s^2}{2} \int^{+\infty}_{-\infty} 
d \nu \, |x|^{1 + 2 i \nu} |x'|^{1-2 i \nu} \frac{1}{\left(\nu^2 + \frac{1}{4}
\right)^2},
\eq
which is identical to Eq. (\ref{repsigmadd2}).

\subsection{Calculation of $\hat{\sigma}_{g d}$ using eikonal methods}
\label{calculsigmagd}

In this section we compute the elementary Born cross-section $\hat{\sigma}_
{\gamma
 d}/k^2$ of the process  
\beq
\label{dgg}
d(\xb)~ g(k) \rightarrow  ~ d(\xb)
\eq
 for a dipole of transverse size $\underline{x}$
 and a soft gluon of virtuality $\underline{k}^2,$ in light-cone gauge. This
 process is illustrated in figure \ref{grapheA}. 
\begin{figure}[htb]
\centering
\epsfysize=6.0cm{\centerline{\epsfbox{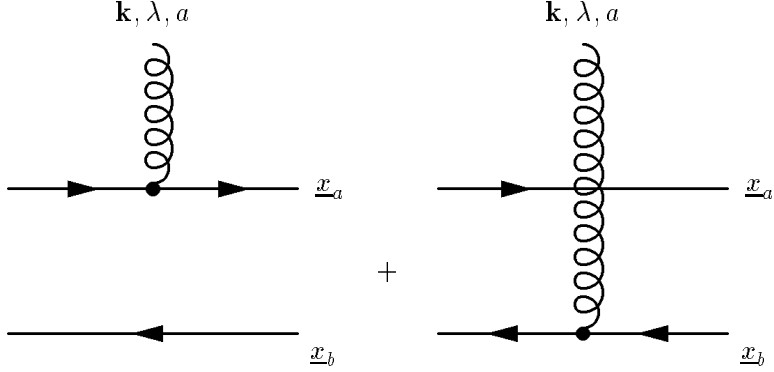}}}
\caption{Amplitudes contributing to the elementary  dipole-gluon 
cross-section.}
\label{grapheA}
\end{figure}
 Consider a dipole of transverse size $\underline{x} = \underline{x}_a - 
\underline{x}_b.$  The corresponding eikonal current has the same expression
 than in covariant gauge (\ref{courant3}). This would not be the case for a
 more
 complicate system for which the current would be rotated in color space (see 
for
 example Ref.\cite{kovchegov}). In light-cone perturbation theory, $k^2 = 2 k^+
 k^- - \underline{k}^2 =0.$ The corresponding current then reads
\beq
\label{courantgd}
j^{a+}(k)= -i g T^a \frac{2 k^+}{\underline{k}^2}\left( e^{-\displaystyle i
\underline{x}_a.\underline{k}} - e^{-\displaystyle i\underline{x}_b.
\underline{k}}
\right).
\eq
The Born cross-section then reads, summing over color and polarization of the
 emitted gluon and averaging over the color of the dipole,
\beqa
\label{born1}
&&\hspace{-1.6CM}\frac{\hat{\sigma}_{g d}}{\kb^2} = \frac{1}{2 (2 \pi)^2N_c} 
Tr|j^+\epsilon^{-\lambda}|^2 \nonumber \\
&& \hspace{-.85CM} = \frac{1}{2 (2 \pi)^2N_c} \sum_{\lambda=1,2} \sum_a g^2 \, 
Tr(T^aT^a) \left(e^{\displaystyle -i\kb.\xb_a} - e^{\displaystyle -i\kb.\xb_b} 
\right) \left((e^{\displaystyle i\kb.\xb_a} - e^{\displaystyle i\kb.\xb_b} 
\right)
4 \left(\frac{\kb.\epsilonb^{\lambda}}{\kb^2}\right)^2.
\eqa
Here one takes into account the fact that the dominant polarization of the 
soft emitted
gluon in light-cone gauge corresponds to 
\beq
\label{polap}
{\epsilon_\mu}^{\lambda} =(\stackrel+0,\stackrel-{\frac{\underline{k}.
\epsilonb^{\lambda}}{k_{+}}},\stackrel\bot\epsilonb^{\lambda})\, \sim
 (\stackrel+0,\stackrel-{\frac{\underline{k}.\epsilonb^{\lambda}}{k_{+}}},
\stackrel\bot0)\,
\eq
in the soft limit.
This finally yields
\beq
\label{born2}
\frac{\hat{\sigma}_{g d}}{\kb^2} = \frac{g^2 C_F}{2 (2 \pi)^2} \left( 2 -
 e^{\displaystyle i\kb.\xb} - e^{\displaystyle -i\kb.\xb} \right)\frac{4}{\kb^2} 
=  \frac{\alpha_s N_c}{\pi} \left( 2 - e^{\displaystyle i\kb.\xb} - e^{
\displaystyle -i\kb.\xb} \right) \frac{1}{\kb^2}.
\eq
Let us make two comments. First, the dominant polarization of the exchanged
gluon is longitudinal, due to the expression (\ref{polap}). However, the
 polarization tensor in light-cone gauge reads
\beq
\label{poltenseur}
d^{\mu\nu} = g^{\mu\nu} - \frac{\eta^{\mu}k^{\nu}+\eta^{\nu}k^{\mu}}{\eta.k}.
\eq
In the Regge limit, since $j_{\nu}$ is proportionnal to $\bar \eta,$ the 
relevant
 component of this tensor is $d^{\mu-},$ and since $\epsilon^+=0$ in light cone
 gauge, it turns out that $\epsilon_{\mu}d^{\mu-}=\epsilon^id^{i-},$ and thus
 the
 exchanged gluon has a physical polarization, as it is expected in high-energy 
factorization.

\noindent
Secondly, when integrating this cross-section over $k$, one can recover the
 dipole
emission kernel which was constructed in Ref. \cite{mueller94}.

\subsection{Calculation of the onium-onium cross section in the laboratory 
frame of one onium}
\label{referentiel}
In this appendix we illustrate the frame invariance of
the onium-onium cross section by an explicit calculation
 in the laboratory 
frame where the left-moving heavy quark-antiquark pair is close to be at rest.
Let us first  consider as in section \ref{resultatexact} the scattering of
two heavy quark-antiquark pairs of transverse sizes $\underline{x}_1$ and
$\underline{x}_2$ which scatter through the exchange of a pair of gluons
between two elementary excited dipoles of transverse sizes $\underline{x}'_1$
and $\underline{x}'_1.$
The corresponding expression for the scattering amplitude $F^{(1)}$ is 
displayed in formula (\ref{calculF1}). Using the same trick which led us
to formula (\ref{calculF2}), we integrate over $d^2 \underline{b}_{int} d^2 
\underline{x}'_1$ and $d^2 \underline{w}_2$, which yields
\beqa
\label{calculF2ref}
&&\hspace{-.5cm} F^{(1)}(\underline{x}_1,\underline{x}_2,\tilde{Y},\underline{b})=
 -2 \frac{\alpha_s^2 (16)^2}{(2 \pi)^6}\frac{\pi^4}{2}
 \sum_{n_1 = -\infty}^{+\infty}  \sum_{n =
 -\infty}^{+\infty} \int^{+\infty}_{-\infty} \frac{d \nu_1}{(2 \pi)^3}
  \int^{+\infty}_{-\infty} d \nu \int
 \frac{d^2\underline{w}_1}{{x'_1}^2}   \nonumber\\
&&\hspace{-.5cm} \times \, \int d^2 \underline{w} \int \frac{d^2
 \underline{x}'_1}{2 \pi {x'_1}^2} \,  \int d^2
 \underline{b}_1 
 \,\left(\nu_1^2 + \frac{n_1^2}{4} \right) \,
 \nonumber \\
&&\hspace{-.5cm}\times \, \left(\nu^2 + \frac{n^2}{4}\right) \,
 \frac{1 + (-1)^n}{\left( \nu^2 + \left(\frac{n-1}{2} \right)^2 \right) 
 \left( \nu^2 +
 \left ( \frac{n+1}{2} \right)^2 \right)} \, \exp
 \left(\frac{2 \alpha_s N_c}{\pi}(\chi(n_1,\nu_1)\tilde{Y}_1
 +\chi(n,\nu)\tilde{Y}_2)\right) \nonumber \\
&&\hspace{-.5cm}\times \, E^{n_1,\nu_1}\left(\underline{b}_1 +
 \frac{\underline{x}'_1}{2} - \underline{w}_1,\underline{b}_1 -
 \frac{\underline{x}'_1}{2} -
 \underline{w}_1\right) \, E^{n_1,\nu_1*}\left(\frac{\underline{x}_1}{2} -
 \underline{w}_1,-
 \frac{\underline{x}_1}{2} - \underline{w}_1\right) \nonumber \\
&&\hspace{-.5cm}\times \,  E^{n,\nu*}\left(\underline{b}_1 +
 \frac{\underline{x}'_1}{2} - \underline{w},\underline{b}_1 -
 \frac{\underline{x}'_1}{2} - \underline{w}\right)
 \, E^{n,\nu}\left(\frac{\underline{x}_2}{2} - \underline{w}+\underline{b},
\frac{-\underline{x}_2}{2} - \underline{w}+\underline{b}\right) 
\eqa
Integrating over $d^2\underline{x}'_1 d^2\underline{b}_1$ and 
$d^2 \underline{w}$ would give the result (\ref{calculF3}). It would show
in particular that in the previous formula $n_1$ can be replaced by $n$
and $\nu_1$ by $\nu$. Since $\tilde{Y} = \tilde{Y}_1 + \tilde{Y}_2,$
the previous formula can be written as
\beq
\label{calculF3ref}
F^{(1)}(\underline{x}_1,\underline{x}_2,\tilde{Y},\underline{b}) = - \frac{1}{2} \int
 d^2
 \underline{b}_1 \int
 \frac{d^2 \underline{x}'_1}{2 \pi {x'_1}^2} n(\underline{x}_1,
\underline{x}'_1,\tilde{Y},\underline{b}_1) \, \sigma_{DD}(\underline{x}'_1,
\underline{x}_2,\underline{b}_1-\underline{b}),
\eq
which is exactly what one would obtain when computing this process in the 
boosted frame where the left-moving onium is close to be at rest.
Indeed, let us consider the dipole distribution inside the left-moving onium
in its laboratory frame, namely $n(\underline{x}_2,\underline{x}'_2,\tilde{Y}=0,
\underline{b}_2).$
Using expression (\ref{distn2}), it reads, using the completeness condition for
the functions $E^{n,\nu}$ (see equation (25) of Ref. \cite{lipatov86})) 
\beqa
\label{distnref}
\hspace{.1cm} n(\underline{x}_2,\underline{x}'_2,\tilde{Y}=0,
\underline{b_2})\nonumber \\
&&\hspace{-4cm} = \sum_{n=-\infty}^{+\infty} 16 \int
 \frac{d \nu}{(2 \pi)^3} \frac{d^2 w}{{x'_2}^2}
 \left(\nu^2 + \frac{n^2}{4} \right)  \,
 E^{n,\nu}\left(\bb_2 + \frac{\xb_2'}{2} - \wb,\bb-\frac{\xb'_2}{2} -
 \wb\right)
 E^{n,\nu*}\left(\frac{\xb_2}{2} - \wb,-\frac{\xb_2}{2} - \wb\right)
\nonumber \\
&&\hspace{-4cm} = \frac{(2 \pi)^4}{(2 \pi)^3} {x_2}^2 \,  \delta^2\left(
\underline{b}_2 +
 \frac{\underline{x}_2' - \underline{x}_2}{2}\right) \, \delta^2 
\left(-\underline{b}_2 + \frac{\underline{x}_2' - \underline{x}_2}{2}\right)
\nonumber \\
&&\hspace{-4cm}= 2 \pi x_2^2 \, \delta^2(\underline{b}_2) \,
 \delta^2(\underline{x}_2' - \underline{x}_2).
\eqa
Calculating the scattering amplitude $F^{(1)}$ with formula (\ref{F1b1})
in this laboratory frame, one gets
\beqa
\label{calculF1b1equivref}
&&\hspace{-1cm}F^{(1)}(x_1,x_2,b,\tilde{Y}) = -\de \int^{\infty}_0 \frac{d^2
 \underline{x}'_1}
{2 \pi {x'_1}^2} \frac{d^2 \underline{x}'_2}{2 \pi {x'_2}^2} d^2
 \underline{b}_1
 \,d^2 \underline{b}_2 \, d^2 (\underline{b}'_2 - \underline{b}'_1) \,
\delta^2(\underline{b}_1 - \underline{b}_2 - \underline{b}'_1 +
 \underline{b}'_2 -
 \underline{b}) \nonumber \\
&&\times \,  n(\underline{x}_1,\underline{x}'_1,\tilde{Y},\underline{b}_1) \, 2 \pi 
x_2^2 \, \delta^2(\underline{b}_2) \,
 \delta^2(\underline{x}_2' - \underline{x}_2) \,
\sigma_{DD}(\underline{x}'_1,
\underline{x}'_2,\underline{b}'_1-\underline{b}'_2)\nonumber \\
&&\hspace{-1cm} = - \frac{1}{2} \int d^2 \underline{b}_1 \int
 \frac{d^2 \underline{x}'_1}{2 \pi {x'_1}^2} n(\underline{x}_1,
\underline{x}'_1,\tilde{Y},\underline{b}_1) \, \sigma_{DD}(\underline{x}'_1,
\underline{x}_2,\underline{b}_1-\underline{b}),
\eqa
where we have performed in the last step the integration with respect to 
$d^2 \underline{x}'_2 \, d^2\underline{b}_2 \, d^2 (\underline{b}'_2 -
\underline{b}'_1).$
This result is identical to Eq. (\ref{calculF3ref}).

\subsection{Approximate calculation of $F^{(1)}$ in the asymptotic regime}
\label{F1approche}

In this appendix we calculate the onium-onium scattering amplitude at fixed
 impact
parameter in the asymptotic regime.  In that case the distributions of
 dipoles $n(\underline{x}_i,\underline{x}'_i,\tilde{Y}_i,\underline{b}_i)$ can be
 approximated  by keeping only 
the contribution $n=0$ to formula (\ref{nq2exact}).
 We show that even in this 
limit it is important to consider the exact expression (\ref{sigmaddb}) of
 the elementary dipole-dipole cross-section. 
In Ref. \cite{muellerpatel}, the onium-onium cross-section at fixed impact
parameter was computed considering the interaction of two elementary
dipoles of transverse size $\underline{x}'_1$ and $\underline{x}'_1,$
situated {\it at the same point} $\underline{b}'_1=\underline{b}'_2,$
interacting through
 the elementary total cross-section 
$\sigma_{DD}(x'_1,x'_2),$ and the dipole distributions were averaged
over angles. Indeed, the elementary cross-section at fixed 
impact parameter
$\sigma_{DD}(\underline{x}'_1,
\underline{x}'_2,\underline{b}'_1-\underline{b}'_2)$ is
 dominated by configurations 
$\underline{b}'_1 \sim \underline{b}'_2,$
since $\sigma_{DD}(\underline{x}'_1,
\underline{x}'_2,\underline{b}'_1-\underline{b}'_2) \sim \frac 
{1}{|\underline{b}'_1 - \underline{b}'_2|^4}$ when 
$|\underline{b}'_1 - \underline{b}'_2|$ is large 
(this can be seen by expanding 
formula (\ref{sigmaddb})).
Thus, it is possible in first approximation to make 
$\underline{b}'_1 = \underline{b}'_2$ and to replace 
$\sigma_{DD}(\underline{x}'_1,
\underline{x}'_2,\underline{b}'_1-\underline{b}'_2)$ by its integral over
 $\underline{b}'_1 - \underline{b}'_2,$ which is equal to 
$\sigma_{DD}(x'_1,x'_2)$ when averaging over angles (see appendix
 \ref{calculsigmaddq}). Note that in the asymptotic regime where
it is possible to keep only the contributions $n_1=n_2 =0$ in the dipole
 distributions, the elementary cross-section will only contribute 
through the quantum number $n=0,$ since this quantum number is
conserved (see Eqs. (\ref{calculF1}-\ref{calculF3}) for example),
and thus the averaging over angles in the elementary cross-section
is automatically implied  (see Eqs. (\ref{sigmaangle}-\ref{resultatequiv})).
However, we will see below that the approximation of integrating the 
elementary dipole-dipole cross section over impact parameter 
is  exact only   when
 calculating the
{\it total} cross-section.
\\

We thus calculate first the cross-section at fixed impact parameter
in these approximations and compare
the result to the exact formula (\ref{relationFn0}). 
\noindent
 Let us consider two heavy quark-antiquark pairs of transverse
size $\underline{x}_1$ and $\underline{x}_2$, which scatter through  the
 exchange of 
a pair of
gluons between two elementary dipoles, respectively of transverse size 
$\underline{x}'_1$ and $\underline{x}'_2$, and at a distance from the center
 of the quark-antiquark
 pair $\underline{b}_1$ and $\underline{b}_2$. In this approximation, the
 corresponding scattering
 amplitude $F^{(1)}$
 given in Eq. (\ref{F1b1}) simplifies in 
\beq
\label{F1b2approx}
F^{(1)}(\underline{x}_1,\underline{x}_2,\tilde{Y},\underline{b}) 
= -\de \int^{\infty}_0 \frac{dx'_1}{x'_1}
 \frac{dx'_2}{x'_2} d^2 \underline{b}_1 \,
 n(\underline{x}_1,\underline{x}'_1,\underline{b}_1,\tilde{Y}_1) \,
 n(\underline{x}_2,\underline{x}'_2,\underline{b} - \underline{b}_1,
\tilde{Y}_2) \,
 \sigma_{DD}(x'_1,x'_2).
\eq
The rapidity $\tilde{Y}_1$ and 
$\tilde{Y}_2$ are such that $\tilde{Y}=\tilde{Y}_1 + \tilde{Y}_2.$
$F^{(1)}$ involves the elementary  dipole-dipole total cross-section,
 which has been evaluated in \cite{muellerpatel}, and which is calculated in
 appendix \ref{calculsigmadd} using
eikonal techniques.
For two dipoles of transverse size $\underline{x}'_1$ and $\underline{x}'_2$,
 the 
elementary forward dipole-dipole cross-section reads (see Eq. (\ref {Atotal4}))
\beq
\label{efficacedd}
\sigma_{DD}(x'_1,x'_2) = 2 \pi \alpha_s^2 \,x^2_< \left[1 + \ln\frac{x_>}{x_<}
\right],
\eq
where
$x_<\,=Min(x'_1,x'_2)$ et $x_> \, = Max(x'_1,x'_2)$.

 We write $F^{(1)}$ given by Eq. (\ref{F1b2approx}) as
\beq
\label{defFnu}
F^{(1)}(x_1,x_2,b,\tilde{Y})= \int^{\infty}_{-\infty} \frac{d \nu}{2 \pi} F_{\nu}^{(1)}
(x_1,x_2,b).
\eq
Using the expression (\ref{repsigmadd}) of the elementary dipole-dipole cross
 section $\sigma_{DD},$ $F_{\nu}^{(1)}$ then reads 
\beqa
\label{F1}
&& \hspace{-1cm} F_{\nu}^{(1)} = -\frac{\pi \alpha_s^2}{2} \frac{1}{\left(
\nu^2 +
 \frac{1}{4}\right)^2} \int^{\infty}_{-\infty} \frac{d \nu_1}{2 \pi} 
\int^{\infty}_{-\infty} \frac{d \nu_2}{2 \pi} \, \exp\left(\frac{2 \alpha_s
 N_c}{\pi}(\chi(0,\nu_1)\tilde{Y}_1 +\chi(0,\nu_2)\tilde{Y}_2)\right) \nonumber \\
&&\hspace{-1cm} \times \, \int \frac{d^2\underline{x}'_1}{2 \pi x^{2'}_1} \int
 \frac{d^2\underline{x}'_2}{2 \pi x^{'2}_2} \int  d^2 \underline{b}_1 \int d^2
 \underline{b}_2 \, \delta^2(\underline{b}-\underline{b_1}-\underline{b_2}) 
 n_{\nu_1}(x_1,x'_1,\underline{b}_1) \, n_{\nu_2}(x_2,x'_2,\underline{b}_2) \,
 (x'_1)^{1+2 i \nu} (x'_2)^{1-2 i \nu}.\nonumber \\
\eqa
Using the representation 
\beq
\label{delta}
\delta^2(\underline{b}-\underline{b}_1 - \underline{b}_2) = \int \frac{d^2
\underline{q}}{(2 \pi)^2} \exp(-i \underline{q}.(\underline{b}-\underline{b}_1-
\underline{b}_2),
\eq
and the expression (\ref{nq2exact}) of 
$n(\underline{x},\underline{x}',\tilde{Y},\underline{b}),$ one gets
\beqa
\label{F2}
&& \hspace{-1cm} F_{\nu}^{(1)} = -\frac{\pi \alpha_s^2}{2} 
 \frac{1}{\left(\nu^2 + \frac{1}{4}\right)^2} \int
 \frac{d^2\underline{q}}{(2 \pi)^2} \int^{\infty}_{-\infty} 
\frac{d \nu_1}{2 \pi} \int^{\infty}_{-\infty} \frac{d \nu_2}{2 \pi}
 \,\exp\left(\frac{2 \alpha_s N_c}{\pi}(\chi(0,\nu_1)\tilde{Y}_1 +\chi(0,\nu_2)\tilde{Y}_2)
\right)\nonumber \\
&& \hspace{-1cm} \times \,  \int \frac{d^2\underline{x}'_1}{2 \pi x^{'2}_1}
 \int \frac{d^2\underline{x}'_2}{2 \pi x^{'2}_2} \, e^{\displaystyle -i
 \underline{q}.\underline{b}} \, \frac{x_1}{x'_1}\, E_q^{0,\nu_1*}(x_1) \,
 E_q^{0,\nu_1}(x'_1) \, \frac{x_2}{x'_2}\, E_q^{0,\nu_2*}(x_2) \,
 E_q^{0,\nu_2}(x'_2) \, (x'_1)^{1+2 i \nu} \, (x'_2)^{1-2 i \nu} \nonumber \\
\eqa
Let us compute the integrations with respect to $x'_i.$
Defining 
\beq
\label{deff}
f_{\nu_i}(\nu,q)= \int \frac{d^d \underline{x}'_i}{2 \pi} \,
 E_q^{0,\nu_i}(x'_i)^{-2+2 i \epsilon_i \nu}
\eq
where $\epsilon_1 =1$ and $\epsilon_2 =-1,$
this function reads, taking into account the representation (\ref{Eq4}) and 
introducing a dimensional regularization $d=2+2 \, \epsilon$ because of the
 divergency when $x'_i \to 0,$
\beqa
\label{f1}
&&\hspace{-1cm} f_{\nu_i}(\nu,q)=
\frac{4 \pi^3}{b_{0,\nu_i}}\left(\frac{q}{2}\right)^{2 i \nu_i}
 \frac{1}{\Gamma^2(\de +i \nu_i)}  \int^1_0 d \alpha_i \,
 (\alpha_i(1-\alpha_i))^{-\de} \nonumber \\
&&\hspace{-1cm} \times \, \frac{\pi^{\epsilon}}{\Gamma(1+\epsilon)}
 \int^{+\infty}_0 d x'_i \, (x'_i)^{-1+2 i \epsilon_i \nu+2 \epsilon} 
K_{-2 i \nu_i} \left(q x'_i \sqrt{\alpha_i(1-\alpha_i)}\right)\, 
J_0\left(q \frac{x'_i}{2}(1-2 \alpha_i)\right),
\eqa
where we have performed the integration with respect to the angle
 $(\underline{q},\underline{x}'_i).$
One can now perform the integration with respect to $x'_i,$ using the formula
(6.576) of  Ref. \cite{Grad}:
\beqa
\label{Grad1}
\int^{\infty}_0 x^{-\lambda} K_{\rho}(a x) J_{\sigma}(bx) \, dx
= \frac{1}{4}\left(\frac{b}{a}\right)^{\sigma} 
\left(\frac{a}{2}\right)^{\lambda-1} \frac{\Gamma\left(\frac{\sigma-\lambda+1
 + \rho}{2}\right) 
\Gamma\left(\frac{\sigma-\lambda+1 - \rho}{2}\right)}{\Gamma(\sigma+1)}
 \nonumber \\
\times {}_2F_1\left(\frac{\sigma-\lambda+1 + \rho}{2},\frac{\sigma-\lambda+1
 - \rho}{2};\sigma+1;-\frac{b^2}{a^2}\right).
\eqa
Here $\sigma=0$, $\rho = -2 i \nu_i$, $b= \frac{q}{2}(1-2 \alpha_i),$
 $a = q\sqrt{\alpha_i(1-\alpha_i)}$ and $-\lambda = 2 i \epsilon_i \nu -1+2
 \epsilon.$
Thus, 
\beqa
\label{calculf2}
&&\hspace{-1.8cm}f_{\nu_i}(\nu,q)= \frac{4 \pi^3}{b_{0,\nu_i}} \, 
\frac{1}{\Gamma^2(\de +i \nu_i)}  \, 
\frac{\pi^{\epsilon}}{\Gamma(1+\epsilon)} \frac{1}{4}  \int^1_0 d \alpha_i 
 (\alpha_i(1-\alpha_i))^{-\de -i \epsilon_i \nu-\epsilon} \left(\frac{q}{2}
\right)^{2 i \nu_i-2 i \epsilon_i \nu-2 \epsilon} \nonumber \\
&& \hspace{-1.8cm} \times \, \Gamma(i \epsilon_i \nu - i \nu_i+ \epsilon)
\Gamma(i \epsilon_i \nu + i \nu_i+\epsilon)\, {}_2F_1\left(i \epsilon_i \nu
 - i \nu_i+\epsilon,i \epsilon_i \nu + i \nu_i+\epsilon;1;-\frac{1}{4}
\frac{(1-2 \alpha_i)^2}{\alpha_i(1-\alpha_i)}\right).
\eqa
One can now perform the integration with respect to $\nu_i.$
We thus define
\beq
\label{defg}
g^i(\nu,x_i,q)= \int^{+\infty}_{-\infty} \frac{d \nu_i}{2 \pi} E_q^{0,\nu_i*}
(x_i) f_{\nu_i}(\nu,q)\, \exp\left(\frac{2 \alpha_s N_c}{\pi}\chi(0,\nu_i)\tilde{Y}_i
\right).
\eq
The analytic structure of the integrand is very simple: it has poles at
 $\nu_i = \pm (\epsilon_i \, \nu -i \epsilon).$ The integration contour can
 be closed 
either in the upper plane or in the lower plane, and give the same
 contribution. For $g^1$ we close the contour in the lower plane so as to
 pick up the residue at $\nu_1 = \nu,$ and for $g^2$ we close the contour 
in the lower plane so as to pick up the residue at $\nu_2 = -\nu.$ 
Since 
\beq
{}_2F_1\left(0,2i \nu ;1;-\frac{1}{4}\frac{(1-2 \alpha_1)^2}
{\alpha_1(1-\alpha_1)}\right)= {}_2F_1\left(-2i \nu ,0;1;-\frac{1}{4}
\frac{(1-2 \alpha_2)^2}{\alpha_2(1-\alpha_2)}\right)=1,
\eq
one gets
\beqa
\label{g1}
g^i(\nu,x_i,q)= \frac{\pi^3}{b_{0,\epsilon_i\nu}} \, \frac{\Gamma(2i
 \epsilon_i \nu)}{\Gamma^2(\de +i \epsilon_i \nu)}  \, \int^1_0 d \alpha_i 
 (\alpha_i(1-\alpha_i))^{-\de -i \epsilon_i \nu} E_q^{0, \epsilon_i \nu*}(x_i)
 \,\exp\left(\frac{2 \alpha_s N_c}{\pi}\chi(0,\nu)\tilde{Y}_i\right),\nonumber \\
\eqa
since $\chi(0,\nu)$ is an even function of $\nu.$
Performing the integration with respect to $\alpha_i$
\beq
\label{integralealpha}
\int^1_0 d \alpha_i  (\alpha_i(1-\alpha_i))^{-\de -i \epsilon_i \nu}= 
\frac{\Gamma^2(-i \epsilon_i \nu + \de)}{\Gamma(-2 i \epsilon_i \nu + 1)},
\eq
$F^{(1)}_{\nu}$ now reads
\beqa
\label{F3}
F^{(1)}_{\nu} &=& -\frac{\pi \alpha_s^2}{2}  \frac{x_1\, x_2}{\left(\nu^2 +
 \frac{1}{4}\right)^2} \int \frac{d^2\underline{q}}{(2 \pi)^2} 
e^{\displaystyle -i \underline{q}.\underline{b}} \, g^1(\nu,x_1,q) \,
 g^2(\nu,x_2,q) \nonumber \\
&=& -\frac{\pi \alpha_s^2}{2}  \frac{x_1 \, x_2 }{\left(\nu^2 + 
\frac{1}{4}\right)^2} \int \frac{d^2\underline{q}}{(2 \pi)^2}
 e^{\displaystyle -i \underline{q}.\underline{b}}  \frac{\pi^6}{b_{0,\nu}
 \,b_{0,-\nu}} \, \frac{\Gamma(2i \nu)}{\Gamma(1+ 2 i \nu)}  \, 
\frac{\Gamma(-2i \nu)}{\Gamma(1 - 2 i \nu)}\nonumber \\
&&\hspace{1cm} \times \,  E_q^{0,  \nu*}(x_1) \,E_q^{0,  \nu*}(x_2)\, 
\exp\left(\frac{2 \alpha_s N_c}{\pi}\chi(0,\nu)\tilde{Y}\right),
\eqa
\noindent
when taking into account that  $\tilde{Y} = \tilde{Y}_1 + \tilde{Y}_2.$ Note that formula (\ref{F3})
could be equivalently obtained from Eq. (\ref{calculf2}) by performing the
 change of variable $\alpha_i \to z_i = (1 -2 \alpha_i)^2$ and using the
 Watson theorem  \cite{Bateman}.
Taking the conjugate of formula (\ref{Eq4}) and then performing the change
 of variable $\alpha = \alpha'-1$, one can verify that 
\beq
\label{propEq}
E^{0,\nu*}_q(\rho) = E^{0,-\nu}_{-q}(\rho)= E^{0,-\nu}_{q}(\rho).
\eq
Using formula (\ref{annu}), one finally gets
\beq
\label{F4}
F^{(1)}_{\nu} = -\frac{\pi \alpha_s^2}{8}  \frac{x_1 \, x_2}{\left(\nu^2 +
 \frac{1}{4}\right)^2} \int \frac{d^2\underline{q}}{(2 \pi)^2} 
e^{\displaystyle -i \underline{q}.\underline{b}}   E_q^{0,  \nu*}(x_1)
 \,E_q^{0,  \nu}(x_2)\exp\left(\frac{2 \alpha_s N_c}{\pi}\chi(0,\nu)\tilde{Y}\right),
\eq
that is
\beq
\label{F5}
F^{(1)}_{\nu} = -\frac{\pi \alpha_s^2}{8 (\nu^2 + \frac{1}{4})^2}\ x_2^2 \,
 n_{\nu}(x_1,x_2,b,\tilde{Y})=-\frac{\pi \alpha_s^2}{8 (\nu^2 + \frac{1}{4})^2}\ x_1^2
 \, n_{\nu}(x_2,x_1,b,\tilde{Y}).
\eq
which differs from the exact result (\ref{relationFn0}) by a factor
 $\frac{1}{2}.$ Thus, the approximation of using the elementary  
dipole-dipole {\it total} cross-section only gives half the correct result. 

One expects that the previous approximations should be
 correct  when calculating the total cross-section.
Indeed, integrating (\ref{F1b2approx}) with respect to $\underline{b}$,
one gets
\beq
\label{Ftotalapprox1}
F^{(1)}(x_1,x_2,\tilde{Y}) = -\de \int^{\infty}_0 \frac{dx'_1}{x'_1}
 \frac{dx'_2}{x'_2} \, n(\underline{x}_1,\underline{x}'_1,\tilde{Y}_1) \,
 n(\underline{x}_2,\underline{x}'_2,\tilde{Y}_2) \, \sigma_{DD}(x'_1,x'_2).
\eq
From formulae (\ref{ntotal1}) and (\ref{repsigmadd}), this yields
\beqa
\label{Ftotalapprox2}
F^{(1)}(x_1,x_2,\tilde{Y}) &=& -\de  \int_{-\infty}^{+\infty} d\nu \, \sum_{n_1=-\infty}
^{+\infty} \int_{-\infty}^{+\infty} \frac{d \nu_1}{2 \pi} \, \sum_{n_2=-\infty}
^{+\infty} \int_{-\infty}^{+\infty} \frac{d \nu_2}{2 \pi} \,  \int \frac{d^2 
\underline{x}'_1}{2 \pi {x'_1}^2} \frac{d^2 \underline{x}'_2}{2 \pi {x'_2}^2} 
\frac{x_1}{x'_1} \left(\frac{x_1^* x_1'}{x_1 {x_1'}^*}\right)^{n_1/2} \left|
\frac{x_1'}{x_1}\right|^{-2 i \nu_1} \,
\frac{x_2}{x'_2} \left(\frac{x_2^* x_2'}{x_2 {x_2'}^*}\right)^{n_2/2} \left|
\frac{x_2'}{x_2}\right|^{-2 i \nu_2} \exp\left(\frac{2 \alpha N_c}{\pi} (
\chi(n_1,\nu_1) \tilde{Y}_1+ \chi(n_2,\nu_2) \tilde{Y}_2) \right) \nonumber \\
&& \times \,   \frac{\alpha_s^2}{2} \, 
 \frac{1}{\left( \nu^2 + \frac{1}{4} \right)^2} 
 |x'_1|^{1-2 i \nu} |x'_2|^{1+2 i \nu}.
\eqa
Integrating with respect to $\underline{x}'_1$ and $\underline{x}'_2$
and using relation (\ref{integredelta}), the only remaining terms are $n_1=n_2=0,$
 due to the conservation of conformal weight, and one gets
\beq
\label{Ftotalapprox3}
F^{(1)}(x_1,x_2,\tilde{Y})= \frac{\pi \alpha_s^2}{2} \int \frac{d \nu}{2 \pi} \frac{1}{
\left( \nu^2 + \frac{1}{4} \right)^2} |x_1|^{1-2 i \nu}\, |x_2|^{1+2 i \nu} \, 
\exp\left(\frac{2 \alpha N_c}{\pi} (\chi(n_1,\nu_1) \tilde{Y}_1+ \chi(n_2,\nu_2) \tilde{Y}_2) 
\right),
\eq
that is
\beq
F^{(1)}_{\{0,\nu\}}(x_1,x_2)= - \frac{\pi \, \alpha_s^2 x_2^2}{4}\frac{1}{\left( 
\nu^2 + \frac{1}{4} \right)^2}\,  n_{\{0,\nu\}}(x_1,x_2),
\eq
which is exactly the result (\ref{relationFntotal0}). 

\subsection{Expression of $E^{0,\nu}_q$ in terms of Bessel functions}
\label{calculE}
In this appendix we prove the expression (\ref{Eq5}) of $E^{0,\nu}_q.$
Starting from equation (\ref{Eq4}) and making the change of variable
 $\alpha = \sin^2 \frac{t}{2},$ the mixed function $E^{0,\nu}_q$ reads
\beq
\label{calculE1}
E^{0,\nu}_q(\rho) = \frac{4 \pi^3}{b_{0,\nu}} \left(\frac{q}{2}\right)^{2 i \nu}
 \frac{1}{\Gamma^2\left(\frac{1}{2} + i \nu \right)} \int^{\pi}_0 e^{\displaystyle
 i
 \frac{\underline{q}.\underline{\rho}}{2} \cos t} K_{-2 i \nu} \left(\frac{q \rho}
{2}
 \sin t\right)\, dt.
\eq
Let us calculate 
\beq
\label{defC}
C =  \int^{\frac{\pi}{2}}_0 e^{\displaystyle i
 \frac{\underline{q}.\underline{\rho}}{2} \cos t} K_{-2 i \nu} \left(\frac{q \rho}
{2}
 \sin t\right)\, dt \,+ \int^{\pi}_{\frac{\pi}{2}} e^{\displaystyle i
 \frac{\underline{q}.\underline{\rho}}{2} \cos t} K_{-2 i \nu} \left(\frac{q \rho}
{2}
 \sin t\right)\, dt.
\eq
Making the change of variable
$t'=\pi-t$ in the second integral, one gets
\beq
\label{calculC1}
C = 2 \int^{\frac{\pi}{2}}_0 \cos \left(
 \frac{\underline{q}.\underline{\rho}}{2} \cos t \right) \,
 K_{-2 i \nu} \left(\frac{q \rho}{2} \sin t\right)\, dt.
\eq
Combining 
\beq
\label{IK}
K_{\mu} (\lambda) = \frac{\pi}{2 \sin (\pi \mu)} [I_{-\mu}(\lambda)-I_{\mu}(
\lambda)]
\eq
and
\beq
\label{IJ}
I_{\mu} (\lambda) = e^{-i \frac{\pi}{2}\mu} J_{\mu}(e^{i \frac{\pi}{2}}\lambda),
\eq
the Bessel function $K$ can be written as
\beq
\label{KJ}
K_{\mu} (\lambda)= \frac{1}{2} \Gamma(\mu) \Gamma(1-\mu) \,
 \left[e^{i \frac{\pi}{2}\mu} J_{-\mu}(e^{i \frac{\pi}{2}}\lambda) -
 e^{-i \frac{\pi}{2}\mu} J_{\mu}(e^{i \frac{\pi}{2}}\lambda)\right].
\eq
$C$ can then be computed using relation (6.688) of Ref. \cite{Grad})
\beqa
\label{intJcos}
&&\int^{\frac{\pi}{2}}_0 J_{\mu}(z \sin t) \, \cos (x \cos t) \, dt
= \frac{\pi}{2} J_{\frac{\mu}{2}}(y_+) \, J_{\frac{\mu}{2}}(y_-),
\nonumber \\
&& \mbox{where } \quad y_{\pm} = \frac{\sqrt {x^2 + z^2} \pm x}{2}.
\eqa
In our case $\mu = -2 i \nu,$ $x =
 \frac{\underline{q}.\underline{\rho}}{2}$ and $z =
 e^{i \frac{\pi}{2}} \frac{q \rho}{2}.$ Defining $\Psi$ the angle $(\underline{q},
\underline{\rho}),$
and computing $y_{\pm},$
\beq
\label{ypm}
y_{\pm} = \pm \frac{q \rho}{4} e^{\pm \displaystyle i \Psi},
\eq
one gets
\beq
\label{calculC2}
C = \frac{\pi}{2} \Gamma(-2 i \nu) \, \Gamma(1+2 i \nu) \left[J_{i \nu}
\left(\frac{q \rho}{4} e^{\displaystyle i \Psi}\right) \, J_{i \nu}
\left(\frac{q \rho}{4} e^{-\displaystyle i \Psi}\right) -J_{-i \nu}
\left(\frac{q \rho}{4} e^{\displaystyle i \Psi}\right) \, J_{-i \nu}
\left(\frac{q \rho}{4} e^{-\displaystyle i \Psi}\right) \right],
\eq
where we have used the relation 
\beq
\label{JJcoupure}
J_{\mu}\left(e^{i \pi} z \right) = e^{i \pi \mu} J_{\mu} (z)
\eq
in order to get rid of the minus sign arising from the $y_-$ contribution.
Combining expressions (\ref{calculE1}) and (\ref{calculC2}), and using the
expression (\ref{bnnu}) for $b_{n,\nu},$ one finally gets
\beq
\label{calculE2}
E^{0,\nu}_q(\rho) =  \left(\frac{q}{2}\right)^{2 i \nu} \, 2^{-2 i \nu}
\Gamma^2(1- i \nu) \,  \left[J_{i \nu}
\left(\frac{q \rho}{4} e^{\displaystyle i \Psi}\right) \, J_{i \nu}
\left(\frac{q \rho}{4} e^{-\displaystyle i \Psi}\right) -J_{-i \nu}
\left(\frac{q \rho}{4} e^{\displaystyle i \Psi}\right) \, J_{-i \nu}
\left(\frac{q \rho}{4} e^{-\displaystyle i \Psi}\right) \right],
\eq
in agreement with the more general result (10) of Ref. \cite{np}.

\subsection{Properties of the three-points correlation functions $E^{n,\nu}$
and $E^{n,\nu}_q$}
\label{proprieteE}
In this appendix we derive various useful mathematical formulae for the 
functions   $E^{n,\nu}$
and $E^{n,\nu}_q.$
Let us first  show that $E^{n,\nu}$ and $E^{-n,-\nu}$ are related by 
the following expression
\beq
\label{lienE}
E^{n,\nu*}(\rho_{10},\rho_{20}) = \frac{b^*_{n,\nu}}{a_{n,\nu}} 
\int d^2 \rho_{0'}
 E^{n,\nu}(\rho_{10'},\rho_{20'}) |\rho_{00'}|^{-2 + 4 i \nu} 
\left(\frac{\rho^*_{0'0}}
{\rho_{0'0}}\right)^n (-1)^n,
\eq
which corrects formula (A.12) of \cite{lipatov86}.
Consider 
\beq
\label{defT}
T = \int d^2 \rho_{0'} E^{n,\nu}(\rho_{10'},\rho_{20'}) 
|\rho_{00'}|^{-2 + 4 i \nu} \left(\frac{\rho^*_{0'0}}
{\rho_{0'0}}\right)^n.
\eq
Using conformal invariance, we can take $\rho_2 \to \infty$ in order to
 simplify the calculation and restore the $\rho_2$ dependence afterwards by
 requiring the correct conformal transformation property.
$T$ then reads
\beqa
\label{T1}
T &=& \int d^2 \rho_{0'} \left(\frac{1}{\rho_{10'}^*}\right)^n |\rho_{10'}|^n 
|\rho_{00'}|^{-2 + 4 i \nu} \left(\frac{\rho_{0'0}^*}{\rho_{00'}}\right)^n
\left|\frac{1}{\rho_{10'}^*}\right|^{1 + 2 i \nu} \nonumber \\
&=& \int d^2 R' \,
 \left(\frac{R' + \frac{\rho}{2}}
{{R'}^* + \frac{\rho^*}{2}}\right)^{\frac{n}{2}} |R'-R|^{-2 + 4 i \nu} \,
 \left(\frac{{R'}^* - R^*}
{R' - R}\right)^n \, \left|\frac{1}{R' + \frac{\rho}{2}}\right|^{1 + 2 i \nu},
\eqa
where $\rho_{10} = R + \frac{\rho}{2},$ $\rho_{20} = R - \frac{\rho}{2},$
$\rho_{10'} = R' + \frac{\rho}{2}$ and $\rho_{20'} = R' - \frac{\rho}{2}.$
After performing the change of variable $R^" = R' - R$ and introducing $R^"=
\left(R + \frac{\rho}{2}\right)z,$ $T$ reads
\beq
\label{T2}
T = \left(\frac{R^* + \frac{\rho^*}{2}}
{R^* + \frac{\rho}{2}}\right)^{\frac{n}{2}} \, |R + \frac{\rho}{2}|^{-1 + 2 i \nu}
 \,K,
\eq
where 
\beq
\label{defK}
K =  \int d^2 z \, \left(\frac{z+1}{z^*+1}\right)^{\frac{n}{2}} |z|^{-2 + 4 i
 \nu} \left(\frac{z^*}{z}\right)^n \, \left|\frac{1}{z+1}\right|^{1+2 i \nu}.
\eq
Using the techniques developped in Ref. \cite{dotsenko}, this integral can
be computed after performing a Wick rotation for the $y$ integration. 
The corresponding replacement $y \to i y e^{-2 i \epsilon}$ reads
$z+z^* = \alpha+ \beta$ and $z-z^* = (\alpha - \beta) e^{-2 i \epsilon}$ with
$d^2 z = dx \, dy = \frac{1}{2}dz \, dz^* = \frac{i}{2} d\alpha \, d \beta.$
Separating the integration in $\alpha$ in three domains, the non-zero 
contribution is obtained for $\alpha \in [-1,0]$ due to the $i \epsilon.$ 
 Closing the integration 
contour for the $\beta$ integration around the singularity $\beta = 0,$ this
yields
\beqa
\label{K1}
K &=& \frac{i}{2} \int d \alpha \, d \beta \frac{(\alpha+1)^{|n|}}
{\left[(\alpha
+1)(\beta+1) + i \epsilon\right]^{\frac{|n|+1}{2} + i \nu}}
 \frac{\beta^{2 |n|}}{(\alpha \beta + i \epsilon)^{|n| + 1 - 2 i \nu}}
\nonumber \\
&=& \int^1_0 d \alpha \, (1-\alpha)^{\frac{|n| - 1}{2} -  i \nu} \,
 \alpha^{-|n|-1 + 2 i \nu} \sin(|n|+1 - 2 i \nu) \, \int^{+\infty}_0 d 
\beta \frac{\beta^{|n|-1 +2 i \nu}}{(\beta+1)^{\frac{|n|+1}{2}+i \nu}},
\eqa
where we have performed the change of variable $\alpha \to - \alpha.$
These integrals lead to $\beta$ functions. After some straightforward 
calculations, one gets
\beq
\label{K2}
K = \frac{\pi}{2} \frac{1}{-i \nu + \frac{|n|}{2}}
 2^{4 i \nu} \frac{\Gamma \left(-i \nu + \frac{|n|+1}{2}\right)}
{\Gamma \left(i \nu + \frac{|n|+1}{2}\right)} \frac{\Gamma \left(i \nu +
 \frac{|n|}{2}\right)}{\Gamma \left(-i \nu + \frac{|n|}{2}\right)} (-1)^n
 = \frac{b_{n,\nu}}{2 \pi^2} (-1)^n = \frac{a_{n,\nu}}{b_{n,\nu}^*} (-1)^n.
\eq
Combining formulae (\ref{lienE},\ref{T2},\ref{K2}) and using
\beq
\label{Econjugue}
E^{n,\nu *}(\rho_{10},\rho_{20}) = \left(\frac{R^* + \frac{\rho^*}{2}}{R +
 \frac{\rho}{2}}\right)^{\frac{n}{2}}
 \left|R+\frac{\rho}{2}\right|^{-1 + 2 i \nu},
\eq
one finally gets formula (\ref{lienE}) after restoring the correct dependence
 in $\rho_2.$

Note that from  Eq. (\ref{lienE}) one can obtain the corresponding
relation between $E_q^{n,\nu *}$ and $E_q^{n,\nu}.$ Indeed, performing the
 Fourier transform of both sides, one gets
\beqa
\label{lienEq1}
E_q^{n,\nu *}(\rho_1 - \rho_2) &=& \frac{2 \pi^2}{b_{n,\nu}^*} \int
 \frac{d^2 \frac{\underline{\rho}_1+\underline{\rho}_2}{2}}{|\rho_1-\rho_2|} 
\, e^{\displaystyle -i \frac{1}{2}(\underline{\rho}_{10}+
\underline{\rho}_{20}).\underline{q}} E^{n,\nu *}(\underline{\rho}_{10},
\underline{\rho}_{20})
 \nonumber \\
&=& \frac{b_{n,\nu}}{a_{n,\nu}} \, \frac{2 \pi^2}{b_{n,\nu}} \int 
\frac{d^2 \frac{\underline{\rho}_1+\underline{\rho}_2}{2}}{|\rho_1-\rho_2|}
 \, e^{\displaystyle -i \frac{1}{2}(\underline{\rho}_{10'}+
\underline{\rho}_{20'}).
\underline{q}} \,  E^{n,\nu}(\underline{\rho}_{10'},\underline{\rho}_{20'}) 
\nonumber \\
&& \hspace{1cm} \times \,  \int d^2\underline{\rho}_{0'} \,
 e^{\displaystyle i(\underline{\rho}_0-\underline{\rho}_{0'}).\underline{q}}
 |\rho_{00'}|^{-2+4 i \nu} \left(\frac{\rho^*_{0'0}}{\rho_{0'0}}\right)^n
 (-1)^n \nonumber \\
&=&\frac{b_{n,\nu}}{a_{n,\nu}} E^{n,\nu}_{-q}(\rho_1-\rho_2) 
\frac{b^*_{n,\nu}}{2 \pi^2} |q|^{-4 i \nu} \left(\frac{q^*}{q}\right)^n
 e^{-i \delta(n,\nu)},
\eqa
where the last integral is computed performing the  integration $d^2 
\underline{\rho}_{0'}=d^2 \underline{\rho}_{0'0}$ with radial coordinates 
 $\rho_{0'0} = |\rho_{0'0}| \exp(i \phi)$ (see the result (A.11) of 
Ref. \cite{lipatov86}).
Using the definition (\ref{Eq1}),
\beqa
\label{EqE-q}
E^{n,\nu}_{-q}(\rho) &=& \frac{2 \pi^2}{b_{n,\nu}} \int 
\frac{d^2 \underline{R}}{|\rho|}e^{\displaystyle-i \qb.\Rb} \, E^{n,\nu}
\left(R+\frac{\rho}{2},R-\frac{\rho}{2} \right)
= \frac{2 \pi^2}{b_{n,\nu}} \int \frac{d^2 \underline{R}}{|\rho|}
e^{i \qb.\Rb}E^{n,\nu}\left(-R+\frac{\rho}{2},-R-\frac{\rho}{2} \right) 
\nonumber \\
&=& \frac{2 \pi^2}{b_{n,\nu}} \int \frac{d^2 \underline{R}}{|\rho|}
e^{\displaystyle i \qb.\Rb} \, E^{n,\nu}\left(R+\frac{\rho}{2},R-\frac{\rho}{2} 
\right) 
= E^{n,\nu}_{q}(\rho),
\eqa
where we have used the fact that 
\beq
\label{Esymetrie}
E^{n,\nu}(\rho_1,\rho_2) = (-1)^n E^{n,\nu}(\rho_2,\rho_1) = 
E^{n,\nu}(-\rho_2,-\rho_1).
\eq
Thus, using formula (\ref{annu}), one finally gets from Eq. (\ref{lienEq1})
\beq
\label{lienEq2}
 E_q^{n,\nu *}(\rho) = |q|^{-4 i \nu} \left(\frac{q^*}{q}\right)^n 
e^{-i \delta(n,\nu)} E_q^{n,\nu}(\rho),
\eq
in agreement with formula (A.15) of Ref. \cite{lipatov86}.
\\

Note that the function $E^{n,\nu}_q$ also possesses the following property
\beqa
\label{conjugueEq}
E^{n,\nu*}_q(\rho) &=& \frac{2 \pi^2}{b_{n,\nu}^*} \int 
\frac{d^2 \underline{R}}{|\rho|}e^{\displaystyle -i \qb.\Rb} 
\, E^{n,\nu*}\left(R+\frac{\rho}{2},R-\frac{\rho}{2} \right) \nonumber \\
&=& \frac{2 \pi^2}{b_{-n,-\nu}} \int \frac{d^2 \underline{R}}{|\rho|}
e^{\displaystyle i \qb.\Rb} \, E^{n,\nu*}
\left(-R+\frac{\rho}{2},-R-\frac{\rho}{2} \right)
 = E^{-n,-\nu}_q(\rho),
\eqa
where we have used the relation
\beq
\label{conjugueE}
E^{n,\nu*}(\rho_1,\rho_2) = E^{-n,-\nu}(-\rho_2,-\rho_1).
\eq

Let us now prove the following orthonormalisation property for the
 three-points
correlation functions $E_q^{n,\nu}$
\beq
\label{orthoEq1}
\int \frac{d^2 \underline{x}'_1}{2 \pi {x'_1}^2} E_q^{n_1,\nu_1} (x'_1)
 E_q^{n,\nu *}(x'_1) = \pi \delta_{n_1,n} \, \delta(\nu_1 - \nu) +  
\pi \delta_{n_1,-n} \, \delta(\nu_1 + \nu) \, \left(\frac{q}{q^*}\right)^{n_1}
 |q^2|^{2 i \nu_1} e^{i \delta(n_1,\nu_1)}.
\eq
Using the definition (\ref{Eq1}), the left-handside reads
\beqa
\label{orthoEq2}
&&\hspace{-1cm}\int \frac{d^2 \underline{x}'_1}{2 \pi {x'_1}^2}
 E_q^{n_1,\nu_1} (x'_1) E_q^{n,\nu *}(x'_1) = \int 
\frac{d^2 \underline{x}'_1}{2 \pi {x'_1}^2} 
\frac{2 \pi^2}{b_{n,\nu}^*} \, \frac{2 \pi^2}{b_{n_1,\nu_1}}\nonumber \\
&&\hspace{-1cm}\times \, \int 
\frac{d^2 \underline{R} \, d^2 \underline{R'}}{{x'_1}^2} \,
 e^{\displaystyle i \underline{q}.(\underline{R} - \underline{R'})} \,
 E^{n_1,\nu_1}\left(R+ \frac{x'_1}{2}, R- \frac{x'_1}{2}\right) \,
 E^{n,\nu *}\left(R'+ \frac{x'_1}{2}, R'- \frac{x'_1}{2}\right).\nonumber \\
\eqa
Performing the changes of variables 
\beq
\label{changevariable2}
\rho_1 - \rho_0 = R + \frac{x'_1}{2}, \quad 
\rho_2 - \rho_0 = R - \frac{x'_1}{2}, \quad \rho_1 - \rho_{0'} = R' +
 \frac{x'_1}{2}, \quad 
\rho_2 - \rho_{0'} = R' - \frac{x'_1}{2},
\eq
and using the equality 
\beq
\label{elementintegration2}
d^2\underline{R} \, d^2\underline{R}' \, d^2\underline{x}'_1 = 
d^2\underline{\rho}_1 \, d^2\underline{\rho}_2 \, d^2\underline{\rho}_{0'},
\eq
one can apply the orthonormalization condition (\ref{ortho}), which yields
\beqa
\label{orthoEq3}
&& \hspace{-1cm} \int \frac{d^2 \underline{x}'_1}{2 \pi {x'_1}^2} 
E_q^{n_1,\nu_1} (x'_1) E_q^{n,\nu *}(x'_1) = \frac{2 \pi^2}{b_{n,\nu}^*} \,
 \frac{2 \pi^2}{b_{n_1,\nu_1}} \frac{1}{2 \pi} \left[a_{n,\nu} \delta_{n_1,n}
 \, \delta(\nu_1-\nu) \int d^2 \underline{\rho}_{0'} \, 
e^{i \underline{q}.\underline{\rho}_{0'0}} \, \delta^2(\rho_{0'0}) \right. 
\nonumber \\
&& \hspace{1cm} \left. + b_{n_1,\nu_1} \delta_{n_1,-n} \, \delta(\nu+ \nu_1)
(-1)^n  \int d^2 \underline{\rho}_{00'} \, |\rho_{00'}|^{-2 - 4 i \nu}
 \left(\frac{\rho_{00'}}{\rho_{00'}^*}\right)^{n_1} \,
 e^{i \underline{q}.\underline{\rho}_{0'}}\right].
\eqa
Using the result (A.11) of Ref. \cite{lipatov86} for the last integral, one 
finally gets the expected result (\ref{orthoEq1}).

\subsection{Calculation of $F^{(1)}$ in the mixed representation}
In this appendix we show how to get the onium-onium scattering amplitude at
fixed impact parameter using the mixed representation for the dipole 
distribution (see Eq.  (\ref{nq2exact}))  and for the elementary  
dipole-dipole cross-section (see Eq. (\ref{devsigmaq})).
Using the Fourier representation of the $\delta$ distribution, formula
 (\ref{F1b1}) reads
\beqa
\label{F1bq1}
&&F^{(1)}(x_1,x_2,\tilde{Y},b) = -\de \int 
\frac{d^2 \underline{x}'_1}{2 \pi {x'_1}^2}
 \frac{d^2 \underline{x}'_2}{2 \pi {x'_2}^2} d^2 \underline{b}_1 \,d^2 
\underline{b}_2 \, d^2 (\underline{b}'_2 - \underline{b}'_1) \,
\int \frac{d^2 \underline{q}}{(2 \pi)^2} \,
 e^{\displaystyle i \underline{q}.(\underline{b}_1 - \underline{b}_2 - 
\underline{b}_1' + \underline{b}_2' - \underline{b})} \nonumber \\
&&\times \,  n(\underline{x}_1,\underline{x}'_1,\underline{b}_1,\tilde{Y}_1) \,
 n(\underline{x}_2,\underline{x}'_2,\underline{b}_2,\tilde{Y}_2) \, \sigma_{DD}
(\underline{x}'_1,
\underline{x}'_2,\underline{b}'_1-\underline{b}'_2) .
\eqa
Combining this expression with formulae (\ref{nq2exact}) and 
(\ref{devsigmaq}), one
gets
\beqa
\label{F1bq2}
&&\hspace{-.7cm} F^{(1)}(x_1,x_2,\tilde{Y},b) = -\de \int \frac{d^2 \underline{q}}
{(2 \pi)^2} \,
 e^{\displaystyle -i \underline{q}.\underline{b}}
 \int \frac{d^2 \underline{x}'_1}{2 \pi {x'_1}^2}
 \frac{d^2 \underline{x}'_2}{2 \pi {x'_2}^2} \, \sum_{n_1=-\infty}^{+\infty}
 \int_{-\infty}^{+\infty} \frac{d \nu_1}{2 \pi} \frac{x_1}{x'_1} \,
 E_q^{n_1,\nu_1 *} (x_1) \,E_q^{n_1,\nu_1 } (x_1') \nonumber \\
&&\hspace{-.7cm} \times \!  \sum_{n_2=-\infty}^{+\infty}
 \int_{-\infty}^{+\infty} \frac{d \nu_2}{2 \pi} \frac{x_2}{x'_2} \, 
E_q^{n_2,\nu_2 *} (x_2) \,E_q^{n_2,\nu_2} (x_2') \, 2 \alpha_s^2  \, 
\frac{x'_1 \, x'_2}{16} \sum_{n=-\infty}^{+\infty} \int_{-\infty}^{+\infty}
 d \nu \,  \frac{1 + (-1)^n}{\left( \nu^2 + \left(\frac{n-1}{2} \right)^2 
\right) \left( \nu^2 + \left ( \frac{n+1}{2} \right)^2 \right)} \nonumber \\
&&\hspace{-.7cm} \times  \, E^{n,\nu*}_q(x'_1) E^{n,\nu}_q(x'_2)  \,   \exp
 \left(\frac{2 \alpha_s^2 N_c}{\pi}(\chi(n_1,\nu_1)\tilde{Y}_1
 +\chi(n,\nu)\tilde{Y}_2)\right)
\eqa
From the property (\ref{conjugueEq}), one has $E_q^{n_2,\nu_2}=
 E_q^{-n_2,-\nu_2*}.$
We can now apply  the orthonormalization condition (\ref{orthoEq1}), which
 yields 
\beqa
\label{F1bq3}
&&\hspace{-.7cm} F^{(1)}(x_1,x_2,\tilde{Y},b) = -\frac{\pi \alpha_s^2 \, x_1 \,
 x_2}{8} \int \frac{d^2 \underline{q}}{(2 \pi)^2} \,
 e^{\displaystyle -i \underline{q}.\underline{b}} \,
\sum_{n=-\infty}^{+\infty} \int_{-\infty}^{+\infty} d \nu \, E^{n,\nu*}_q(x_1)
 E^{n,\nu}_q(x_2)   \nonumber \\
&& \times \,  \frac{1 + (-1)^n}{\left( \nu^2 + \left(\frac{n-1}{2} \right)^2 
\right) \left( \nu^2 + \left ( \frac{n+1}{2} \right)^2 \right)} \exp
 \left(\frac{2 \alpha_s N_c}{\pi}(\chi(n_1,\nu_1)\tilde{Y}_1
 +\chi(n,\nu)\tilde{Y}_2)\right).
\eqa
From the expression of the dipole density (\ref{nq2exact}),
this finally reads
\beq
\label{F1bq4}
F^{(1)}_{\{n,\nu\}}(\underline{x}_1,\underline{x}_2,\underline{b})= 
-\frac{\pi \alpha_s^2 x_2^2}{8}  \frac{1 + (-1)^n}{\left( \nu^2 +
 \left(\frac{n-1}{2} \right)^2 \right) \left( \nu^2 + \left ( \frac{n+1}{2}
 \right)^2
 \right)} \, n_{\{n,\nu\}}(\underline{x}_1,\underline{x}_2,\underline{b}).
\eq
which is identical to Eq. (\ref{relationFn}) as expected.

\end{appendix}
\eject
\noindent{\large {\bf References}}

\end{document}